\documentclass[twocolumn, twocolappendix]{aastex701}

\usepackage{CJK}
\usepackage{amsmath}	
\usepackage{amssymb}	
\usepackage{enumitem}
\usepackage{verbatim} 
\usepackage[caption=false]{subfig}
\usepackage{array}

\usepackage{listings}
\usepackage{hyperref}

\begin{document}
\begin{CJK*}{UTF8}{gbsn}

\title{Detecting White Dwarf Exoplanets in the Roman Era}


\newcommand{\todo}[1]{\textcolor{red}{ToDo: #1}}
\newcommand{\tday}{T_{\rm day}}
\newcommand{\tdaymax}{T_{\rm day,max}}
\newcommand{\tirr}{T_{\rm irr}}
\newcommand{\tirrroche}{T_{\rm irr,\,Roche}}
\newcommand{\deltar}{\Delta\mathcal{R}}

\author[0000-0003-0525-9647]{Zifan Lin (林梓帆)}
\affiliation{Department of Physics and McDonnell Center for the Space Sciences, Washington University, St. Louis, MO 63130, USA}
\email{lzifan@wustl.edu}

\author[0000-0002-6939-9211]{Tansu Daylan}
\affiliation{Department of Physics and McDonnell Center for the Space Sciences, Washington University, St. Louis, MO 63130, USA}
\email{tansu@wustl.edu}


\begin{abstract}

White dwarfs (WDs) provide an unparalleled opportunity for detecting and characterizing exoplanets because of the large planet-to-star size ratio. However, only one transiting WD exoplanet is confirmed to date, and occurrence rate constraints remain loose because the rapid transits and weak phase curve modulations of WD exoplanets require high-precision, high-cadence photometry. The Galactic Bulge Time Domain Survey (GBTDS) of the Nancy Grace Roman Space Telescope will provide exactly this by observing thousands of WDs at 12.1-minute cadence. The expected WD exoplanet yield of this survey is not yet quantified. Here, we predict the GBTDS WD exoplanet yield over its first three high-cadence seasons, through both transiting and non-transiting (phase curve) channels. We run Monte Carlo detection probability simulations over a grid of planet radius, orbital period, and host properties for a Besan\c{c}on synthetic population of WDs. Jupiter-sized planets dominate the transit yield, with a period-averaged effective survey size of $\overline{N}_\mathrm{eff}\approx8$, requiring an underlying occurrence rate of $\overline{\eta}_1\approx12\%$ for one expected detection, followed by $\overline{N}_\mathrm{eff}\approx2.6$ ($\overline{\eta}_1\approx38\%$) for Neptune-sized and $\overline{N}_\mathrm{eff}\approx1.8$ ($\overline{\eta}_1\approx55\%$) for sub-Neptune-sized planets. Source blending is the major observational challenge in the crowded GBTDS fields, removing $\approx95\%$ of the raw predicted yield. Phase curve provides marginal detections, adding at most $\approx4\%$ to the transit yield. A null detection by GBTDS would constrain $\eta_\mathrm{WD}$ to $<28\%$ for Jupiter-sized planets at $95\%$ confidence. GBTDS will deliver the first WD exoplanet occurrence-rate constraints beyond the solar neighborhood and identify prime targets for future characterization.

\end{abstract}



\section{Introduction} \label{sec:intro}

Following the detection and characterization of the first transiting white dwarf (WD) exoplanet WD 1856+534 b \citep{vanderburg_giant_2020, limbach_thermal_2025, macdonald_aerosols_2026}, the prospect of discovering and studying more exoplanets around WDs is emerging. WDs are the compact remnants of post-main-sequence stellar evolution. A roughly solar-mass WD is so dense that it has an approximately Earth-like radius. Large planet-to-star size ratios and hence large transit depths make terrestrial WD exoplanets readily detectable, in contrast to their counterparts around main-sequence stars with much shallower transits \citep{agol_transit_2011}. Large transit depths also make transiting WD exoplanets favorable targets for atmospheric characterization and biosignature gas search \citep{loeb_detecting_2013, kozakis_high-resolution_2020, kaltenegger_white_2020, lin_h_2022}. Even though only one transiting WD exoplanet has been confirmed to date, indirect evidence suggests that planetary materials around WDs are common. Roughly a quarter to a half of young DA WDs show metal-polluted atmospheres, and many host debris disks \citep[e.g.,][]{zuckerman_ancient_2010, koester_frequency_2014}, implying that planetesimals and even intact planets may be common around WDs.

However, despite extensive searches, WD exoplanets remain rare. In addition to WD 1856+534 b, only a handful of non-transiting WD companions have been detected or tentatively identified. Microlensing observations have revealed intact planets orbiting WDs \citep[e.g.,][]{blackman_jovian_2021, zhang_earth-mass_2024}. \cite{mullally_jwst_2024} reported two directly imaged giant planet candidates around WD 1202-232 and WD 2105-82 with JWST/MIRI, but a second epoch of imaging revealed that neither source shares common proper motion with its host, suggesting that both are background galaxies \citep{mullally_follow-up_2026}. An unresolved infrared excess around WD 2105-82 persists in the second epoch of observation, but whether it arises from a planet or a debris disk remains unknown. The scarcity of WD exoplanets is consistent with existing occurrence rate constraints, which are upper limits from null detections. Using $1{,}148$ WDs observed by K2, \cite{van_sluijs_occurrence_2018} constrained the occurrence of hot Jupiters to $<1.5\%$ and of habitable-zone Earth-sized planets to $<28\%$, broadly consistent with earlier WASP and Pan-STARRS limits \citep{faedi_detection_2011, fulton_search_2014}. Recent population synthesis by \cite{mauch-soriano_predicted_2026} predicted that fewer than $\sim3\pm1.5\%$ of WDs host a planet, predominantly wide-orbit gas giants, broadly consistent with previous surveys.

The lack of tight occurrence-rate constraints stems from the fact that WD exoplanet searches to date are fundamentally limited to the bright, local WD population they can reach. TESS has been searching for WD exoplanets, using both longer cadence modes and its fast 20 s cadence that is short enough to resolve the rapid, roughly a few minutes long transits of WD planets, but only previously known irregular transits by circumstellar debris have been recovered \citep{robert_frequency_2024}. The same limit also applies to the Gaia WD census \citep[e.g.,][]{gentilefusillo_catalogue_2021}, where only bright and nearby bodies are characterized. Ground-based wide-field surveys can reach fainter, more distant targets. The Rubin Observatory's LSST will monitor millions of WDs across half the sky. Simulations predict that LSST could uncover tens to many hundreds of WD planets, depending on the underlying occurrence rate \citep{lund_transiting_2018, cortes_detectability_2019}. LSST sampling is sporadic, however, so these detections will only gradually accumulate over the 10-year survey baseline.

The Nancy Grace Roman Space Telescope (hereafter Roman), officially scheduled to launch on August 30th, 2026, will offer an exciting new opportunity for detecting exoplanets orbiting WDs. Among the three Roman core community surveys, we expect that the Galactic Bulge Time-Domain Survey (GBTDS) will transform the landscape of WD exoplanet detection. Roman's unprecedented combination of high near-infrared sensitivity and wide field of view will let it probe deep into the Galactic bulge and center, extending the exoplanet detection horizon well beyond the solar neighborhood. \cite{wilson_transiting_2023} estimated that the GBTDS will be sensitive to transiting planets across the far side of the bulge, at distances of $\gtrsim$16--20\, kpc. The high-cadence GBTDS seasons will cycle through five bulge fields and one Galactic center field \citep{zasowski_roman_2025}, observing each in the \texttt{F146} band every 12.1\, minutes, which is essential for sampling brief WD exoplanet transits. GBTDS will also offer a long survey baseline in addition to the short cadence. Six high-cadence seasons will span $\approx5$ years, increasing the sensitivity to long-period planets. Together, these capabilities position Roman -- and specifically GBTDS -- as a powerhouse for WD exoplanet detection throughout its mission lifetime.

Several studies have predicted Roman's transiting exoplanet yield, but all of them assume main-sequence hosts. \cite{wilson_transiting_2023} predicted between $\sim$60{,}000 and $\sim$200{,}000 transiting exoplanets in total, among which $\sim$7{,}000--12{,}000 are small ($R_p < 4\,R_\oplus$). Focusing on the faint, late-type end of main-sequence hosts, \cite{tamburo_predicting_2023} predicted that Roman will find $1347^{+208}_{-124}$ small transiting planets around mid-M and ultracool dwarfs, including $\sim$13 terrestrial planets in the habitable zone. These prediction studies excluded WDs. This work presents the first predicted yields for both transiting and non-transiting WD exoplanets from the Roman GBTDS.

This paper is structured as follows. Section~\ref{sec:method} outlines our methods. Sections~\ref{sec:results_transit} and \ref{sec:results_nontransit} present our predictions for the yields of transiting and non-transiting WD exoplanets with GBTDS, respectively. We discuss the implications, caveats, and potential future directions of this work in Section~\ref{sec:discussion}. Finally, we summarize our conclusions in Section~\ref{sec:conclusions}.

\section{Methods} \label{sec:method}
Here, we describe the methods we use to predict the GBTDS yield of WD exoplanets. We perform a Monte Carlo (MC) simulation over a large grid of system parameters, which returns, for each cell, a detection signal-to-noise ratio (SNR) and a detection probability ($P_{\rm det}$) for transiting and non-transiting planets. The MC grid requires several inputs, which we introduce in turn. We first assemble a library of WD spectral energy distributions (SEDs) in Section~\ref{subsec:sed}. We then build a synthetic GBTDS WD population (Section~\ref{subsec:population}). By feeding the SEDs into \texttt{Pandeia} \citep{pontoppidan_pandeia_2016}, we compute the per-exposure Roman \texttt{F146} photometric precision (Section~\ref{subsec:noise}). Given the noise model, we define the SNR and $P_{\rm det}$ grid (Section~\ref{subsec:grid}) used to simulate three types of planetary signals: the primary transit (Section~\ref{subsec:transit_model}), secondary eclipses, and periodic phase curves (Section~\ref{sec:nontransit_model}). We then calculate the Roche distance to exclude planets on tidally unstable orbits (Section~\ref{subsec:roche}), and simulate two important observational effects for the crowded GBTDS fields: source blending and dust extinction (Section~\ref{subsec:obs_effects}). Finally, we perform an injection-recovery test using \texttt{allesfitter} \citep{gunther_allesfitter_2021} to validate that the predicted SNR and $P_{\rm det}$ are still robust for a blind search (Section~\ref{subsec:injection}).

\subsection{White Dwarf Spectral Energy Distribution Library} \label{subsec:sed}
We compile a library of WD SEDs as inputs for the subsequent photometric simulations. We adopt the DA model atmosphere grids from \cite{koester_white_2010} for $T_\mathrm{eff} = 7000$--$80{,}000$\,K models and supplement them with cool WD SEDs for $T_\mathrm{eff} = 4000$, $5000$, and $6000$\,K, identical to those used by \cite{kozakis_uv_2018, kozakis_high-resolution_2020} and were computed as described in \cite{saumon_near-uv_2014}. We assume $\log g = 8.0$ for all WDs, because this value is near the peak of the known WD population, according to the Montreal White Dwarf Database (MWDD; \citealt{dufour_montreal_2017}). The input WD SEDs are processed into a common format (wavelength in nm and flux in $\mathrm{W\,m^{-2}\,nm^{-1}}$) and are read at runtime as input to the \texttt{Pandeia} noise model (Section~\ref{subsec:noise}). 
The processed library and other data products of this work are publicly available on Zenodo \citep{lin_detecting_2026}.

\subsection{Synthetic White Dwarf Population in the GBTDS Fields}
\label{subsec:population}
\begin{figure}[t]
    \centering
    \includegraphics[width=0.95\columnwidth]{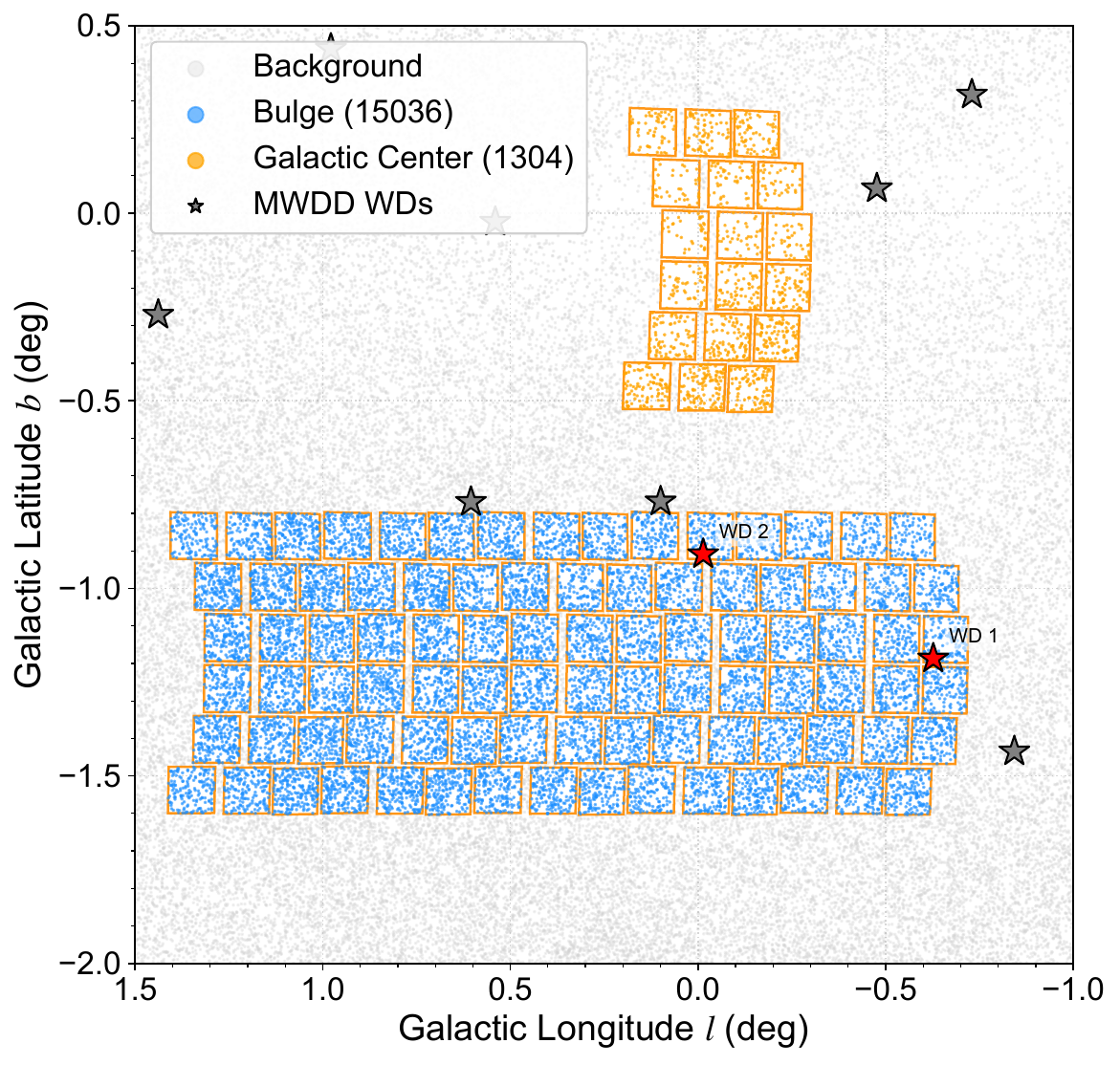}
    \caption{Besan\c{c}on synthetic WD population overplotted with the GBTDS footprints. WDs are color coded by their locations: (blue) in the five Galactic bulge fields, (orange) in the Galactic center field, and (gray) in the background. Known WDs from MWDD are shown as star symbols, with the two WDs in GBTDS fields highlighted in red. WD1 = Gaia DR3 4056053113067931520. WD2 = Gaia DR3 4057210997558861312.}
    \label{fig:besancon_gaia_in_GBTDS_fields}
\end{figure}

Predicting the GBTDS yield requires a realistic census of the WDs that Roman will observe. A synthetic population is necessary because Roman will probe deep into the Galactic bulge and center, while known WD populations to date primarily cover the solar neighborhood \citep[e.g.,][]{gentilefusillo_catalogue_2021, obrien_40_2024} and are therefore severely incomplete at Galactic distances. We queried the MWDD and found only two confirmed WDs, both detected by Gaia, within the six GBTDS fields (Gaia DR3 4056053113067931520 and Gaia DR3 4057210997558861312), at 0.13 kpc and 0.22 kpc away, respectively. While these two WDs can serve as nearby, bright WD sources for validation, they do not represent the WD population that Roman will observe (Figure \ref{fig:besancon_gaia_in_GBTDS_fields}).

\begin{figure*}[t]
    \centering
    \includegraphics[width=\textwidth]{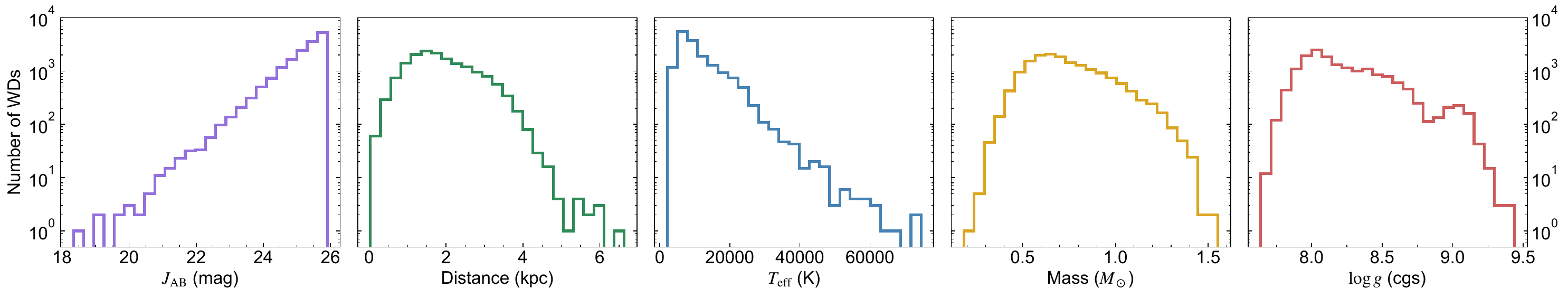}
    \caption{Distributions of in-field WD host properties from the Besan\c{c}on population synthesis. Five parameters are shown from left to right: source magnitude $J_{\rm AB}$ (as a proxy of Roman \texttt{F146} brightness), distance, effective temperature ($T_{\rm eff}$), mass, and surface gravity $\log g$. WD counts in all histograms are shown on a log scale.}
    \label{fig:besancon_wd_population_histogram}
\end{figure*}

We simulate the GBTDS field WD population using the Besan\c{c}on Galaxy model \citep{robin_synthetic_2003, czekaj_besancgalaxy_2014}, version \texttt{m1612}. Besan\c{c}on is a population synthesis model that predicts the stellar content along any Galactic line of sight, including four populations: the thin disk, thick disk, the stellar halo, and the outer bulge. We treat WDs as a separate, self-consistent population. All WDs in Besan\c{c}on are assumed to be DA type using the cooling tracks and atmosphere models of \cite{bergeron_photometric_1995}, which is consistent with our SED database assumption (Section \ref{subsec:sed}), and are complemented by the \cite{chabrier_is_1999} model at cool temperatures. We query a cone toward the GBTDS pointings, centered at Galactic coordinates $(\ell, b) = (0.25^\circ, -0.75^\circ)$ with a solid angle of $6.25\,\mathrm{deg}^2$ and a heliocentric distance out to $10$\, kpc.

Photometry for the population is computed using the SDSS$+$JHK system, and we impose a single magnitude limit in the $J$ band, $J_\mathrm{Vega} \le 25$, leaving all other bands unconstrained. Here, the $J$ band serves as a proxy for Roman's \texttt{F146} filter, which covers 0.927--2\,$\mu$m and encompasses the $J$ band, because Besan\c{c}on does not include the Roman photometric system. As a precedent, \cite{tamburo_predicting_2023} also selected their Besan\c{c}on sample using near-infrared ($r, i,z,J$) magnitude cuts as \texttt{F146} proxies. We convert to AB magnitudes using $J_\mathrm{AB} = J_\mathrm{Vega} + 0.91$, where the $J$-band offset comes from \cite{blanton_k-corrections_2007}, so the synthetic population reaches $J_\mathrm{AB} = 25.91$. This is a conservative completeness limit. \cite{tamburo_predicting_2023} limited their sample to cool main sequence stars with \texttt{F146} $<21$, while \cite{wilson_transiting_2023} considered \texttt{F146} $<23$.

Dust extinction is included in the Besan\c{c}on model by default, following the 3D dust map of \cite{marshall_modelling_2006}, so the catalog magnitudes, colors, and $V$-band extinction magnitude $A_V$ already include reddening. Therefore, we do not apply a dust correction in the noise model or elsewhere to avoid double counting (see Section \ref{subsubsec:dust} for details).  

Our Besan\c{c}on query returns $53{,}669$ WDs over the full $6.25\,\mathrm{deg}^2$ region, which we mask by Roman Wide Field Instrument (WFI) pointings to filter out those invisible to GBTDS. Shape and exact coordinates of the WFI pointings were adopted from the \texttt{pysiaf} package \citep{shannon_osborne_spacetelescopepysiaf_2026}, which contains 18 detector apertures per GBTDS field. A WD is counted as in-field if it falls within any detector aperture of any pointing. In total, there are $16{,}340$ in-field WDs ($15{,}036$ in the bulge fields and $1{,}304$ in the Galactic-center field). Figure \ref{fig:besancon_gaia_in_GBTDS_fields} shows the Besan\c{c}on synthetic population within the GBTDS pointings. In addition to brightness, the Besan\c{c}on model also returns the distance, effective temperature, mass, and surface gravity of the synthetic WDs. The distributions of host properties are shown in Figure \ref{fig:besancon_wd_population_histogram}.

\subsection{\texttt{Pandeia} Single-Exposure Noise Model and Seasonal Light Curve}
\label{subsec:noise}

The detectability of a planetary signal in its host light curve ultimately depends on the photometric precision of each exposure. We compute the per-exposure $1\sigma$ photometric uncertainty $\sigma_F$ using the \texttt{Pandeia} \citep{pontoppidan_pandeia_2016} exposure-time calculator, with engine version \texttt{2026.1} for Roman. We run \texttt{Pandeia} for WD hosts spanning a range of $T_{\rm eff}$ and $J_{\rm AB}$ to acquire noise amplitude, and inject Gaussian noise of this level into the simulated light curves. 
We configure \texttt{Pandeia} to reproduce the GBTDS high-cadence setup exactly, using WFI imaging in the \texttt{F146} filter, the \texttt{im\_66\_6} MultiAccum table, and a single $66$\,s exposure per epoch. In each 72-day high-cadence season, Roman will take 8390 such exposures at a 12.1-minute cadence, and the first three high-cadence seasons are separated by roughly 6-month intervals, spanning a $\approx436$-day baseline \citep{zasowski_roman_2025}. 

While six high-cadence GBTDS seasons are planned in total, we simulate only the first three seasons to be performed in the first $\sim$2 years of the mission, leaving the final three seasons, which will not occur until 2030, to future work once Roman's in-flight performance is constrained.

We model each WD as a point source at the center of the scene and set the \texttt{Pandeia} extinction to zero, because dust extinction is applied upstream to the Besan\c{c}on simulation and is therefore not double-counted here (see Section~\ref{subsubsec:dust}). For each WD, we pass the template SED corresponding to its $T_\mathrm{eff}$ from our library (Section~\ref{subsec:sed}) to set the spectral shape, and normalize it to the WD's apparent $J$-band magnitude, $J_\mathrm{AB}$, from the Besan\c{c}on population synthesis. Because we assume the SED shape depends only on $T_\mathrm{eff}$ (with fixed DA atmospheres and $\log g=8.0$, see Section \ref{subsec:sed}) while the brightness is set by $J_\mathrm{AB}$, $T_\mathrm{eff}$ and brightness are two independent axes of our simulation (see details in Section \ref{subsec:grid}). \texttt{Pandeia} returns the extracted source count rate $F$ and the per-exposure signal-to-noise ratio $\mathrm{SNR}_1$, from which we define the $1\sigma$ photometric uncertainty $\sigma_F$ as
\begin{equation}
    \sigma_F = F / \mathrm{SNR}_1 .
\end{equation}
We evaluate $\sigma_F$ once for each unique $(T_\mathrm{eff}, J_\mathrm{AB})$ pair and use it as the width of the noise distribution for that host throughout. A new noise value is then drawn from this distribution at each point of the light curve, independently for every MC trial and orbital configuration.

\subsection{Detection Probability Grid}
\label{subsec:grid}

To predict the GBTDS WD exoplanet yield, we evaluate detectability across a grid of relevant planet and host parameters, assigning each grid cell a detection SNR and a detection probability $P_\mathrm{det}$. In Section~\ref{sec:results_transit} and \ref{sec:results_nontransit}, we fold $P_\mathrm{det}$ over a realistic WD population to obtain the expected number of transit or phase curve detections.

\begin{deluxetable*}{lllcl}
\tabletypesize{\scriptsize}
\tablewidth{0pt}
\tablecaption{Planet system parameter grid for detection probability simulation. \label{tab:grid}}
\tablehead{
\colhead{Parameter} & \colhead{Symbol} & \colhead{Values} & \colhead{$N$} & \colhead{Notes}
}
\startdata
WD brightness          & $J_\mathrm{AB}$  & $18.0$--$24.5$                                   & 13 & --          \\
Orbital period         & $P$              & 500 min to 581 d                                 & 40 & Log-spaced  \\
Eccentricity           & $e$              & $\{0,\,0.2,\,0.4,\,0.6,\,0.8\}$                  & 5  & --          \\
Argument of periastron & $\omega$         & $\{0,\,90,\,180,\,270\}^\circ$                   & 4  & --          \\
Planet radius          & $R_p$            & $\{0.5,\,1,\,2,\,3,\,4,\,11\}\,R_\oplus$         & 6  & --          \\
WD temperature         & $T_\mathrm{eff}$ & $\{5,\,7,\,10,\,15,\,20\}\times10^3$\,K          & 5  & --          \\
WD mass                & $M_\mathrm{WD}$  & $\{0.4,\,0.6,\,0.8,\,1.0,\,1.1,\,1.2\}\,M_\odot$ & 6  & --          \\
H envelope             & --               & Thick                                            & 1  & --          \\
\cutinhead{Channel-specific extension}
WD brightness       & $J_\mathrm{AB}$  & $\{25.0,\,25.5,\,26.0\}$ & 3 & Faint WD extension (transit only) \\
WD temperature      & $T_\mathrm{eff}$ & $\{25000,\,30000\}$      & 2 & Hot WD extension (transit only)   \\
H envelope          & --               & Thin                     & 1 & Transit only                      \\
Impact parameter    & $b$              & $\{0,\,0.3,\,0.6,\,0.9\}$ & 4 & Transit \& eclipse only          \\
Orbital inclination & $i$              & $\{20,\,80\}^\circ$       & 2 & Phase curve only                  \\
\enddata
\end{deluxetable*}

The base grid for the detection probability simulation (Table~\ref{tab:grid}) is defined as follows. The transit, eclipse, and phase curve channels all share this base grid. We assume 13 host magnitudes between $J_{\rm AB}=18$ and 24.5, 40 orbital periods between 500 min (roughly the minimum period of a rocky exoplanet allowed by the Roche limit, see Section~\ref{subsec:roche}) and 581 days, 5 eccentricities, 4 arguments of periastron, 6 planetary radii from Mars-sized to Jupiter-sized, 5 WD effective temperatures from cool (5000 K) to hot (20,000 K), and 6 WD masses from 0.4 to $1.2\, M_\odot$. The base grid assumes that all WDs have a thick hydrogen envelope \citep{bedard_spectral_2020}. In total, the base grid contains $13\times40\times5\times4\times6\times5\times6 = 1{,}872{,}000$ cells.

Note that, for the orbital period axis, we extend beyond the $\approx436$-day total duration of the first three high-cadence GBTDS seasons. Exoplanets at the long-period end of the $P$ axis transit at most once or twice within the first three seasons, insufficient for a confident periodic detection. The Kepler pipeline requires $\geq3$ transits to robustly constrain the period and reduce false positives \citep{burke_planet_2017}. Nonetheless, long-period systems are worth simulating. Single- and double-transit events are essential for detecting long-period planets in surveys with limited baselines, and two widely separated transits can already restrict the period to a discrete set of aliases that future observations can resolve \citep[e.g.,][]{cooke_resolving_2021, hawthorn_tess_2024}. A marginal long-period candidate flagged in the first three seasons can therefore be revisited in the later three seasons to confirm its periodicity through additional transits. The phase curve channel, moreover, does not rely on discrete transit events, although its sensitivity declines as the orbital period approaches or exceeds the observing baseline.

For edge-on systems detectable via transit or eclipse, we further assume four impact parameters $b = \{0,\,0.3,\,0.6,\,0.9\}$. For non-transiting planets that require phase-curve detection, the impact parameter is irrelevant. Instead, we define two orbital inclinations: $i = 20^\circ$ and $i = 80^\circ$. To study the influence of envelope thickness on the detection SNR, we additionally simulate a thin H envelope case for the transit grid, holding all other grid parameters fixed. We further generate two extension grids for faint and hot WDs for the transit channel (Table~\ref{tab:grid}).

Note that the WD radius ($R_{\rm WD}$), which is essential for simulating the transit light curve, is not among the free grid parameters. This is because $R_{\rm WD}$ can be computed from the theoretical evolutionary sequences of \cite{bedard_spectral_2020} if $M_\mathrm{WD}$, $T_\mathrm{eff}$, and the H envelope thickness are known. 

We compute detection SNR and $P_\mathrm{det}$ for each cell using MC simulation. In each trial, we draw a random orbital phase, generate the corresponding synthetic light curve (Section~\ref{subsec:noise}), and attempt to recover the planetary signal. We use a variable number of MC trials per cell, assigning more trials to longer-period planets. This treatment is necessary because the GBTDS samples at a fixed $12.1$-min cadence, so whether a brief WD transit happens to fall in an exposure window varies sharply with orbital phase. Averaging over more random-phase trials smooths out this aliasing.

Here, we define the exoplanet detection SNRs reported by the model grid. Let $s_i$ be the noise-free transit, eclipse, or phase curve signal from an analytical model, and $d_i = s_i + n_i$ the same signal with added noise. The expected detection SNR is \citep{kipping_snr_2023}
\begin{equation}
   \mathrm{SNR}_\mathrm{exp} = \sqrt{\sum_i \frac{s_i^2}{\sigma_F^2}},
\end{equation}
which is an upper limit assuming that we know the noise-free signal exactly. Taking noise into consideration, we will get a different measured SNR
\begin{equation}
   \mathrm{SNR}_\mathrm{meas} = \frac{1}{\mathrm{SNR}_\mathrm{exp}} \sum_i \frac{s_i d_i}{\sigma_F^2},
\end{equation}
which is derived from the $\chi^2$-based SNR definition of \cite{kipping_snr_2023}. In the limit of $d_i = s_i$ (i.e., noise $n_i=0$), $\mathrm{SNR}_\mathrm{meas}$ falls back to $\mathrm{SNR}_\mathrm{exp}$. We classify a MC trial as a detection if $\mathrm{SNR}_\mathrm{meas} \ge 7$, following the conventional $7\sigma$ cutoff \citep{jenkins_tests_2002}, and define $P_\mathrm{det}$ as the fraction of recovered trials for that specific grid point.

\subsection{Transit Signal}
\label{subsec:transit_model}

We model the transit light curve with the analytic occultation model of \cite{mandel_analytic_2002}, using the Keplerian sky-projected star-planet separation defined in \cite{winn_transits_2014}, which can handle the edge case of total occultation due to the large planet-to-star size contrast.

We treat the WD disk as uniform and neglect limb darkening. Because WD exoplanet transits are typically deep or even total, the transit depth is dominated by the fraction covered rather than the surface brightness profile. Hence, limb darkening has a negligible effect on the detection significance and the predicted yield. In addition, limb darkening weakens toward longer wavelengths, as shown in the WD limb darkening coefficient tables for LSST $ugrizy$ bands \citep{gianninas_limb-darkening_2013}, so the effects in the \texttt{F146} band should be minimal. Limb darkening does, however, affect the ingress and egress shapes of the light curve, which matters when we constrain planetary parameters from a blind fit. We therefore include limb darkening in the \texttt{allesfitter} injection-recovery test (Section~\ref{subsec:injection}), adopting coefficients from \cite{gianninas_limb-darkening_2013}.

\subsection{Phase Curve and Secondary Eclipse Signals}
\label{sec:nontransit_model}

Transit is not the only way to detect exoplanets orbiting WDs. Non-transiting planets reflect incoming radiation and emit in the infrared, producing a sinusoidal phase curve signal. If the WD host is sufficiently hot and emits mainly in the visible, while the planet is irradiated enough to produce a strong thermal signal, we may be able to detect a WD exoplanet via its phase curve in the Roman \texttt{F146} band. In addition, secondary eclipses produce a signal directly related to the planet's dayside emission temperature, which may corroborate transit detections. Eclipse and phase curve observations have been performed for many highly irradiated planets orbiting main-sequence stars \citep{lin_persistent_2026}, and the same methodology can be readily expanded to WD systems. Roman's near-infrared sensitivity will open an era of detailed study of planetary thermal emission for exoplanets in WD systems and main-sequence systems alike.

To simulate planetary flux for the phase curve and eclipse signals, we compute the planet's dayside and nightside temperatures using the energy balance model of \cite{cowan_statistics_2011}. The equilibrium temperature at the planet's substellar point is $T_0 = T_\mathrm{eff}\sqrt{R_\mathrm{WD}/a}$. The dayside temperature, parametrized by the Bond albedo $A_B$ and the heat-redistribution efficiency $\varepsilon$, is
\begin{equation}
    T_\mathrm{day} = T_0\,(1 - A_B)^{1/4}\left(\frac{2}{3} - \frac{5}{12}\varepsilon\right)^{1/4},
\end{equation}
and the nightside temperature is
\begin{equation}
   T_\mathrm{night} = T_0\,(1 - A_B)^{1/4}\left(\frac{\varepsilon}{4}\right)^{1/4},
\end{equation}
where $\varepsilon = 0$ implies no heat transport ($T_\mathrm{night} = 0$) and $\varepsilon = 1$ gives full redistribution.

At a given orbital phase $\phi$ ($\phi = 0$ at transit, $\phi = \pi$ at secondary eclipse), the infrared emission from the planet is a linear combination of its dayside and nightside contributions. Writing the phase angle as $\cos\alpha = -\sin i \cos\phi$, where $i$ is the orbital inclination, the thermal flux is weighted by the two hemispheres by their projected visible fraction, $f_\mathrm{day} = \tfrac{1}{2}(1 + \cos\alpha)$ and $f_\mathrm{night} = \tfrac{1}{2}(1 - \cos\alpha)$,
\begin{equation}
   F_\mathrm{thermal}(\phi) = \left(\frac{R_p}{R_\mathrm{WD}}\right)^2 \frac{f_\mathrm{day}\,B(T_\mathrm{day}) + f_\mathrm{night}\,B(T_\mathrm{night})}{B(T_\mathrm{eff})},
\end{equation}
where $B(T)$ is the blackbody radiance integrated over the \texttt{F146} bandpass.

Additionally, we calculate the contribution to the phase curve from the reflected flux. For simplicity, we treat the planet as a Lambertian sphere, for which the geometric albedo is $A_g = \tfrac{2}{3}A_B$. The reflected flux is therefore \citep[see e.g.,][]{robinson_inferring_2026}
\begin{equation}
   F_\mathrm{refl}(\phi) = \frac{2}{3}\,A_B \left(\frac{R_p}{a}\right)^2 \Phi(\alpha),
\end{equation}
where $\Phi(\alpha)$ is the Lambert phase function defined as
\begin{equation}
   \Phi(\alpha) = \frac{\sin\alpha + (\pi - \alpha)\cos\alpha}{\pi}.
\end{equation}

We assign $A_B$ and $\varepsilon$ values based on planetary radius. Planets below the $1.6\, R_\oplus$ radius valley \citep{fulton_california-kepler_2017} are assumed to be bare rocks with no heat redistribution ($A_B = 0$, $\varepsilon = 0$). In contrast, planets above the radius valley are assumed to have values consistent with volatile envelopes ($A_B = 0.2$, $\varepsilon = 0.3$, following \citealt{cowan_statistics_2011}).

\subsubsection{Phase Curve}
\label{subsubsec:phasecurve}
The planet's combined thermal and reflected flux, $F_\mathrm{total}(\phi) = F_\mathrm{thermal}(\phi) + F_\mathrm{refl}(\phi)$, varies periodically over each orbit as its dayside rotates into and out of view. The signal $s_i$ we search for is therefore not a sudden drop in brightness, but the periodic variation itself, $s_i = F_\mathrm{total}(\phi) - \overline{F_\mathrm{total}}$, where $\overline{F_\mathrm{total}}$ is the flux averaged over a full orbit. In the simulation grid for SNR and $P_\mathrm{det}$, we replace the impact-parameter axis with orbital inclination, simulating two values, $i = 20^\circ$ and $80^\circ$. The advantage of the phase curve is that it does not require edge-on geometry, which matters more for WDs than for main-sequence hosts because the small size of a WD implies a low transit probability. Moreover, phase curve detection does not rely on short-duration events like transits and eclipses, so it is feasible with lower-cadence data, potentially extending our pipeline to high-latitude sources.

\subsubsection{Secondary Eclipse}
\label{subsubsec:eclipse}
Secondary eclipses are modeled by combining the planetary flux model discussed above with the same analytic occultation geometry as the primary transit \citep{mandel_analytic_2002, winn_transits_2014}, with the foreground and background bodies swapped. Unlike eclipses in main-sequence systems, planets are often comparable to or larger than the WD host, so it is likely that a planet's unocculted dayside is still visible during an eclipse. The eclipse signal $s_i$ is therefore the deviation of $[1 - f_\mathrm{obs}(\phi)]\, F_\mathrm{total}(\phi)$ from the out-of-eclipse baseline, where $f_\mathrm{obs}(\phi)$ is the fraction of the planet's disk covered by the WD. Because eclipsing systems are edge-on and therefore also transit, we evaluate secondary eclipses on the same base grid as transits (Table \ref{tab:grid}) and discuss the contribution by the eclipse signals to the survey yield (Section~\ref{subsec:eclipse_null}).

\subsection{Roche-Limit Cut}
\label{subsec:roche}
Our simulation grid (Section \ref{subsec:grid}) extends to very short (500 min) orbital periods, which roughly corresponds to the orbital period of a planet that stays in the WD habitable zone for billions of years \citep[$a=0.0085$ AU, following][]{kozakis_uv_2018}. At this extremely close orbital separation, we need to calculate the Roche limiting distance to ensure the planet is not tidally disrupted, so our grid parameters remain physical.

We adopt the incompressible fluid body Roche limit \citep{rappaport_roche_2013}, which will become relevant for volatile-rich planets with low bulk densities,
\begin{equation}
   P_\mathrm{Roche} \simeq \sqrt{\frac{3\pi(2.44)^3}{G\,\rho_p}},
\end{equation}
which is a function of $\rho_p$, the planet's bulk density, alone.

We only use $R_p$ as a grid parameter, but do not assume planet mass. Even though statistical mass-radius relations exist \citep{chen_probabilistic_2016}, exoplanets have very diverse interior compositions, making it challenging to assign a bulk density based only on $R_p$. Therefore, we broadly categorize our synthetic planets into three groups and assign them densities based on solar system planets. Small ($0.5\, R_\oplus$ and $1\, R_\oplus$) planets are assumed to have an Earth-like density of 5514 kg m$^{-3}$. Intermediate (2--$4\, R_\oplus$) planets are assumed to have a Neptune-like density of 1638 kg m$^{-3}$. Large ($11\, R_\oplus$) planets are assumed to have a Jupiter-like density of 1326 kg m$^{-3}$. Based on these bulk densities, we derive the $P_{\rm Roche}$ of small, intermediate, and large planets to be 0.223 d (321 min), 0.410 d (590 min), and 0.455 d (655 min), respectively, implying that some of our shortest period grid points are unphysical for volatile-rich planets. We apply a cut when reporting the estimated yield by removing planets with $P < P_{\rm Roche}$. 

\subsection{Observational Effects} \label{subsec:obs_effects}
So far, our simulation grid has treated each WD as an isolated point source. The actual GBTDS fields, however, are highly crowded and suffer from severe dust extinctions. Here, we discuss how source blending and dust extinction affect planet-yield predictions.

\subsubsection{Source Blending}
\label{subsubsec:blending}
In the crowded GBTDS fields, the photometric aperture around a target WD will inevitably contain light from unresolved neighboring stars. To quantify this dilution, we define the blending factor $D$ as
\begin{equation}
   D = \frac{F_\mathrm{WD}}{F_\mathrm{WD} + F_\mathrm{bg}},
\end{equation}
where $F_\mathrm{WD} = 10^{-0.4\, J_\mathrm{AB}}$ is the WD flux and $F_\mathrm{bg}$ is the integrated flux of all unresolved sources within a circular 3-pixel Roman WFI aperture. Given that WFI has $0.11''$ pixels \citep{zasowski_roman_2025}, the total angular radius is $0.33''$.

\begin{figure}[t]
    \centering
    \includegraphics[width=0.95\columnwidth]{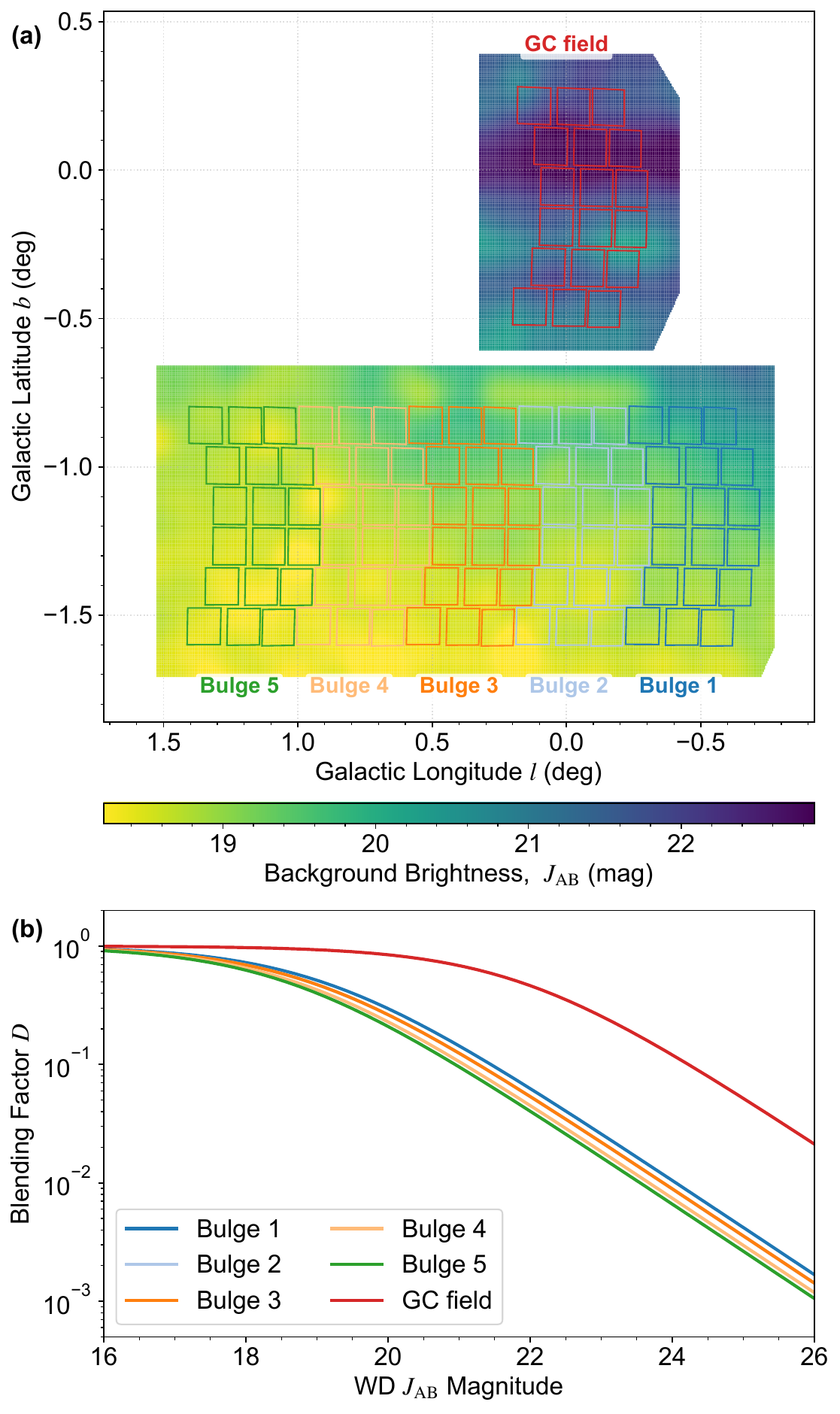}
    \caption{(a) Background brightness map in GBTDS fields. Roman WFI footprints are shown and color coded by field. (b) Blending factor $D$ as a function of WD source magnitude in the five bulge fields and the Galactic center (GC) field.}
    \label{fig:source_blending}
\end{figure}

We build a map of the background flux $F_\mathrm{bg}$ as a function of sky position in and around the GBTDS fields. We sample the GBTDS footprint on a dense $(\ell, b)$ grid with 1472 locations. At each location, we query the Besan\c{c}on model for a small cone of solid angle $2\times10^{-4}\,\mathrm{deg}^2$ out to $10$\, kpc in the same SDSS$+$JHK photometry as the synthetic WD catalog (Section \ref{subsec:population}), for all stellar types and all magnitudes. 
At each grid location, we integrate the $J$-band fluxes of all synthetic stars and normalize by the ratio of the $0.33''$ aperture area to the cone solid angle. These per-location values are interpolated to produce a 2D background flux map (Figure \ref{fig:source_blending}a). Then, we assign each synthetic WD in the GBTDS fields a blending factor $D$ using the equation above to quantify how much the planetary signal is diluted.

The resulting blending factor depends strongly on both sky position and WD brightness. In the bulge fields, a relatively bright WD at $J_\mathrm{AB} = 18$ retains roughly half of its flux ($D \approx 0.4$), but $D$ falls to $\sim$0.1 by $J_\mathrm{AB} = 20$, $\sim$0.02 by $J_\mathrm{AB} = 22$, and $\sim3\times10^{-3}$ by $J_\mathrm{AB} = 24$ (Figure \ref{fig:source_blending}b). The blending factor is higher (less dilution) in the Galactic center field than in the bulge fields. A WD at $J_\mathrm{AB} = 18$ in the Galactic center field has $D \approx 0.94$, and $D$ remains $\sim$0.7 at $J_\mathrm{AB} = 20$, $\sim$0.3 at $J_\mathrm{AB} = 22$, and $\sim$0.06 at $J_\mathrm{AB} = 24$. However, this is not a genuine reduction in crowding, but a dust extinction effect. The Besan\c{c}on background magnitudes already include interstellar extinction from the \cite{marshall_modelling_2006} 3D dust model, and the severe dust extinction toward the Galactic center dims the background stars, lowering $F_\mathrm{bg}$ and hence raising $D$. The target WD, however, is dimmed by dust extinction too, so the apparent gain in $D$ does not translate into a better signal. This makes dust extinction the next observational effect we must quantify.

\subsubsection{Dust Extinction} \label{subsubsec:dust}

\begin{figure*}[t]
    \centering
    \includegraphics[width=\textwidth]{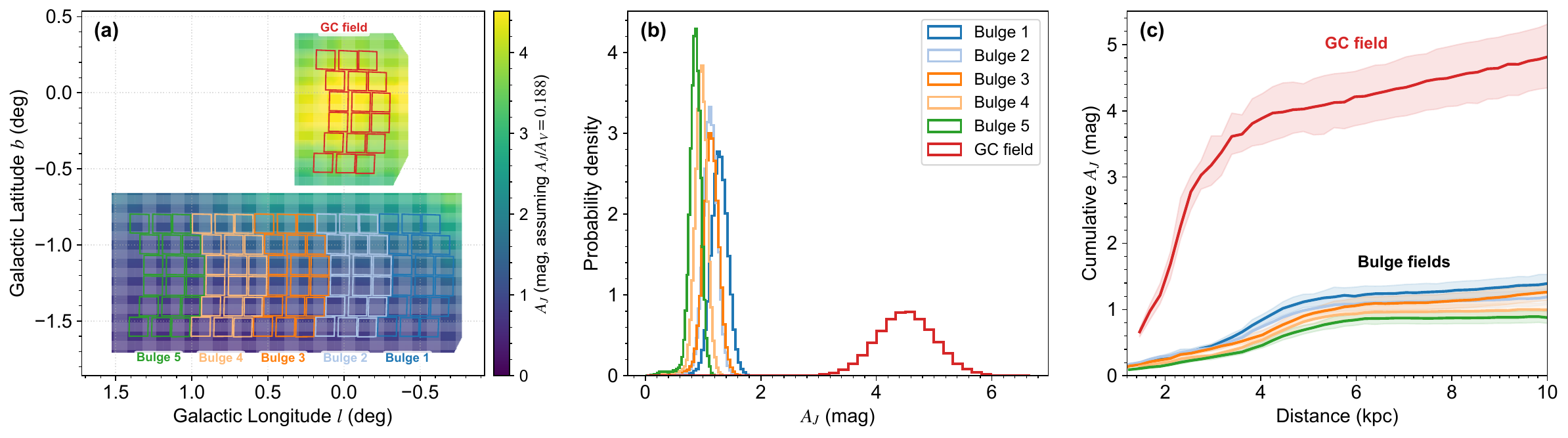}
    \caption{(a) $J$-band dust extinction map in the GBTDS fields. Roman WFI footprints are shown and color coded by field. (b) Probability density distribution of extinction $A_J$ in the five GBTDS bulge fields and the Galactic center (GC) field. (c) Cumulative $A_J$ as a function of heliocentric distance. Dust extinction is moderate for the Galactic bulge but severe for the Galactic center, with $A_J$ distribution peaks at $\sim4.5\,\mathrm{mag}$ (a source is $\sim60\times$ fainter due to dust extinction).}
    \label{fig:dust_extinction}
\end{figure*}

Interstellar dust extinction remains a major challenge for Roman surveys, both dimming and reddening sources. Dust extinction is particularly severe towards the Galactic bulge and center, and can build up steeply with distance along the line of sight. Our Besan\c{c}on population synthesis already includes a 3D dust extinction model by \cite{marshall_modelling_2006}, so the WD brightnesses that set the photometric noise already account for dust attenuation. We therefore do not re-apply extinction, and instead display the dust map as a diagnostic of conditions across the GBTDS fields.

For this purpose, we convert the per-star $V$-band extinction $A_V$ reported by the Besan\c{c}on catalog into the $J$ band. Using OGLE observations in the Galactic bulge, \cite{nishiyama_interstellar_2008} reported the extinction ratio to be $A_J/A_V = 0.188\pm0.005$, which we adopt here for the conversion. In Figure \ref{fig:dust_extinction}, we show the $J$-band dimming due to dust extinction as a function of Galactic coordinate $(\ell, b)$, the $A_J$ probability density distribution in each GBTDS field, and the cumulative $A_J$ as a function of distance along the GBTDS line of sight. In the five bulge fields, thanks to Roman's infrared capability, the attenuation is limited: $A_J \lesssim 1\,\mathrm{mag}$ out to $\sim4$ kpc and $\lesssim 2\,\mathrm{mag}$ out to $\sim10$ kpc. In the Galactic center field, dust extinction is much more severe, with $A_J$ peaking around $\sim$$4.5\,\mathrm{mag}$, which means the source appears $\sim60\times$ fainter.

\subsection{Injection-Recovery Tests}
\label{subsec:injection}
So far, our prediction grid relies on a matched filter method where we know the true signal exactly. Such prior knowledge is not available for a real survey. To validate whether a blind search without any prior knowledge can still yield a similar level of SNR and detection probability as predicted by the simulation grid, we perform an independent injection-recovery test using \texttt{allesfitter} \citep{gunther_allesfitter_2021}.

We generate a single three-season Roman light curve spanning $\approx436$ days, the total span of three 72-day seasons separated by $\sim6$ months from start to start, with \texttt{Pandeia}-derived per-exposure noise (Section~\ref{subsec:noise}), and inject transits into it. This test targets validation of the detection grid rather than a full reproduction, so we vary only the two most relevant parameters for the transit signal -- orbital period and planet radius -- while keeping all other parameters fixed. Orbital period is spaced logarithmically from $500$\,min to $218$\,d (40 values). Note that the $P$ upper bound is lower than that of the SNR grid in Section~\ref{subsec:grid} because we require $\geq3$ transits for a robust detection following the Kepler convention to reduce the false alarm rate \citep{burke_planet_2017}. We assume the same six planet radii used in the SNR grid, $R_p \in \{0.5, 1, 2, 3, 4, 11\}\, R_\oplus$. The host is fixed at a thick-envelope WD with $M_\mathrm{WD} = 0.6\,M_\odot$ and $R_\mathrm{WD} = 0.0128\,R_\odot$ \citep[following typical values from][]{kozakis_uv_2018}, $T_\mathrm{eff} = 7000$\,K, and $J_\mathrm{AB} = 20.0$. Unlike the detection grid, the injected transits include quadratic limb darkening, with coefficients from \cite{gianninas_limb-darkening_2013} for the LSST $y$-band as a proxy of Roman \texttt{F146}. We inject each of the $240$ cells $15$ times with independent phase and noise realizations ($3600$ injections in total) to produce a smooth recovery fraction per cell.

Each injected planet is searched with the transit least squares method \citep{hippke_optimized_2019}.
We consider an injection recovered if the recovered period matches the injected one to within $1\%$ and the signal detection efficiency (SDE) $\ge 7$. The per-cell recovery fraction is defined to be $N/15$, where $N$ is the number of successful recoveries among the 15 injections.

We present a comparison between the simulation grid and the injection-recovery test in Section~\ref{subsec:injection} below. The two methods are not expected to be numerically identical because the injection-recovery test incurs penalties from a blind search, but broad agreement confirms the validity of our simulation grid.

\section{Transiting Planet Detection Yields} \label{sec:results_transit}

Here, we present our predicted yields of transiting WD exoplanets for the GBTDS. Previous works with similar objectives \citep{montet_measuring_2017, wilson_transiting_2023, tamburo_predicting_2023} predicted $N_{\rm det}$, the total number of transiting planets Roman is expected to detect. Reporting such a number is possible for main-sequence hosts because earlier surveys have well constrained their planet occurrence rates ($\eta$) \citep[e.g.,][]{hsu_occurrence_2019}. For WDs, however, only upper bounds exist on the planet occurrence rate ($\eta_\mathrm{WD}$) due to the lack of transit detections beyond WD 1856+534 b. We therefore report two quantities in place of $N_{\rm det}$: $\overline{N}_\mathrm{eff}$ and $\overline{\eta}_1$. The effective survey size $N_\mathrm{eff}(R_p, P)$ is the expected number of detections if every WD hosts exactly one planet ($\eta_\mathrm{WD} = 100\%$) of radius $R_p$ at period $P$. Its period average $\overline{N}_\mathrm{eff}(R_p)$, calculated assuming a log-uniform period prior, is the corresponding quantity marginalized over orbital period. The required occurrence rate $\overline{\eta}_1(R_p) = 1/\overline{N}_\mathrm{eff}$ is the intrinsic WD exoplanet occurrence rate for which one detection is expected for GBTDS. We define these quantities in detail in Section~\ref{subsec:yield_framework}.

\subsection{Period-Averaged Roman GBTDS Transit Yield Prediction}
\label{subsec:results_yield}

\begin{figure}[t]
    \centering
    \includegraphics[width=\columnwidth]{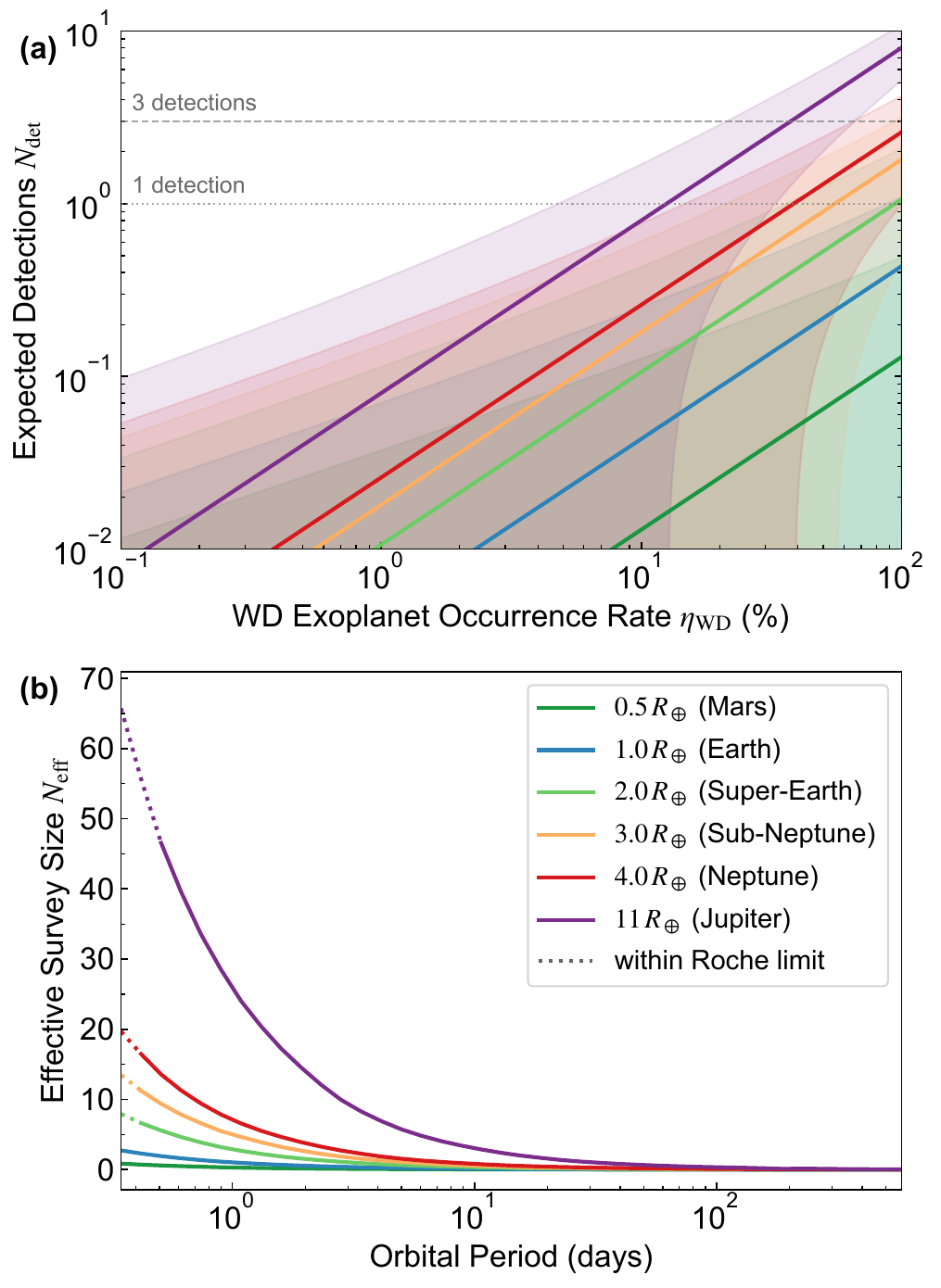}
    \caption{(a) Expected number of transiting WD exoplanet detections by Roman GBTDS as a function of the underlying WD exoplanet occurrence rate, $\eta_{\rm WD}$, color coded by planet size. 
    (see also Table \ref{tab:transit_yield}). The shaded areas show $\pm1\sigma$ Poisson uncertainties. (b) $N_{\rm eff}$ as a function of orbital period. Periods that are too short and within the Roche limit are demarcated as dotted curves.}
    \label{fig:expected_det_vs_eta_WD}
\end{figure}

Over the first three high-cadence seasons, Jupiter-sized ($11\, R_\oplus$) planets dominate the expected detections from the GBTDS, with $\overline{N}_{\rm eff}\approx8$, while Neptune-sized ($4\, R_\oplus$) planets follow with $\overline{N}_{\rm eff}\approx2.6$. Sub-Neptune-, Super-Earth-, Earth-, and Mars-sized planets have decreasing $\overline{N}_{\rm eff}$ of $\approx1.80$, $\approx1.06$, $\approx0.43$, and $\approx0.13$, respectively (Table \ref{tab:transit_yield}). In terms of the underlying occurrence rate required to get one expected detection from the first three GBTDS seasons, the six planet types have the following $\overline{\eta}_1$: Jupiter-sized planets have $\overline{\eta}_1 \approx 12\%$, Neptune-sized planets have $\overline{\eta}_1 \approx 38\%$, Sub-Neptune-sized planets have $\overline{\eta}_1 \approx 55\%$, while Super-Earth-sized, Earth-sized, and Mars-sized planets have $\overline{\eta}_1 \approx94\%$, 230\%, and 771\%, respectively. Figure \ref{fig:expected_det_vs_eta_WD} shows the expected number of transiting WD exoplanet detections by Roman GBTDS as a function of $\eta_{\rm WD}$ (panel a) and $N_{\rm eff}$ as a function of planetary orbital period (panel b).

To assess the effects of realistic observational constraints on the transit yield, we further report two values: the raw value directly from the prediction grid with no filters applied, and the ``Roche only'' value with only unphysical planets within Roche limits removed with no source blending applied. 

Source blending is the most severe limiting factor for transiting WD exoplanet detection in the crowded GBTDS fields. Applying source blending removes the vast majority of detections across all planet sizes ($\approx$95\% for Jupiter and $\approx$97--98\% for Neptune and smaller, see Table \ref{tab:transit_yield}). This is because faint sources dominate the WD host population. The $16{,}340$ in-field WDs have a median brightness of $J_\mathrm{AB} \approx 25.4$, and only $\sim$2\% are brighter than $J_\mathrm{AB} = 23$ (Figure \ref{fig:besancon_wd_population_histogram}). In the extremely crowded bulge fields, planetary signals around these faint WDs are diluted by a factor of $D \sim 10^{-2}$ to $10^{-3}$ (Figure \ref{fig:source_blending}). 

\begin{deluxetable}{lcccc}
\tabletypesize{\scriptsize}
\tablewidth{\columnwidth}
\tablecaption{Effective survey size marginalized over log-uniform period bins $\overline{N}_\mathrm{eff}$ for transiting WD exoplanets in the GBTDS, by planet size, for the raw, Roche limit, and the final Roche plus source blending tiers. The final column gives $\overline{\eta}_1 = 1/\overline{N}_\mathrm{eff}$, the occurrence rate required for one expected detection. \label{tab:transit_yield}} 
\tablehead{
\colhead{Planet size} & \colhead{$\overline{N}_\mathrm{eff}$ raw} & \colhead{$+$Roche} & \colhead{$+$blend} & \colhead{$\overline{\eta}_1$}
}
\startdata
Mars (0.5\,$R_\oplus$)        & 5.0   & 5.0   & 0.13  & 771\% \\
Earth (1\,$R_\oplus$)         & 26.9  & 26.9  & 0.43  & 230\% \\
Super-Earth (2\,$R_\oplus$)   & 53.8  & 48.3  & 1.06  & 94\%  \\
Sub-Neptune (3\,$R_\oplus$)   & 72.2  & 65.0  & 1.80  & 55\%  \\
Neptune (4\,$R_\oplus$)       & 89.8  & 81.0  & 2.60  & 38\%  \\
Jupiter (11\,$R_\oplus$)      & 211.1 & 172.1 & 8.01  & 12\%  \\
\enddata
\end{deluxetable}

In contrast, the Roche limit cut, which removes planets on tidally unstable orbits too close to the WD (Section~\ref{subsec:roche}), has a relatively minor and density-dependent effect on the transit yield. Because we assign a single bulk density to each of the three size classes, the cut affects each class differently. Large ($11\, R_\oplus$, Jupiter-density) planets are the most affected, losing $\approx$18\% of their yield ($\overline{N}_\mathrm{eff}$ drops from $211.1$ to $172.1$), because the low bulk density places the Roche limit at the longest period and removes the most close-in giant planets, which are also the most easily detectable. Intermediate ($2$--$4\,R_\oplus$, Neptune-density) planets lose $\approx$10\%. Small ($0.5$--$1\, R_\oplus$, Earth-density) planets are unaffected, because their high bulk density places the Roche limit below the shortest period sampled by our grid.

\subsection{Effective Survey Size Across the System Parameter Space}
\label{subsec:neff_resolved}

The $\overline{N}_\mathrm{eff}$ values and the corresponding $\overline{\eta}_1$ reported in Section \ref{subsec:results_yield} are marginalized over a log-uniform period prior. These values offer a conservative estimate of the overall Roman GBTDS yield as a function of planet size, but they do not show how $N_\mathrm{eff}$ depends on other system parameters. We discuss the orbital period, host brightness, and eccentricity dependence of $N_\mathrm{eff}$ in Sections \ref{subsubsec:period_dependence}, \ref{subsubsec:brightness_dependence}, and \ref{subsubsec:ecc_dependence}, respectively. In Section \ref{subsubsec:favorable_wds}, we describe the characteristics of the most favorable WD systems for the Roman GBTDS exoplanet search.

\subsubsection{Orbital Period Dependence}
\label{subsubsec:period_dependence}

\begin{figure}[t]
    \centering
    \includegraphics[width=\columnwidth]{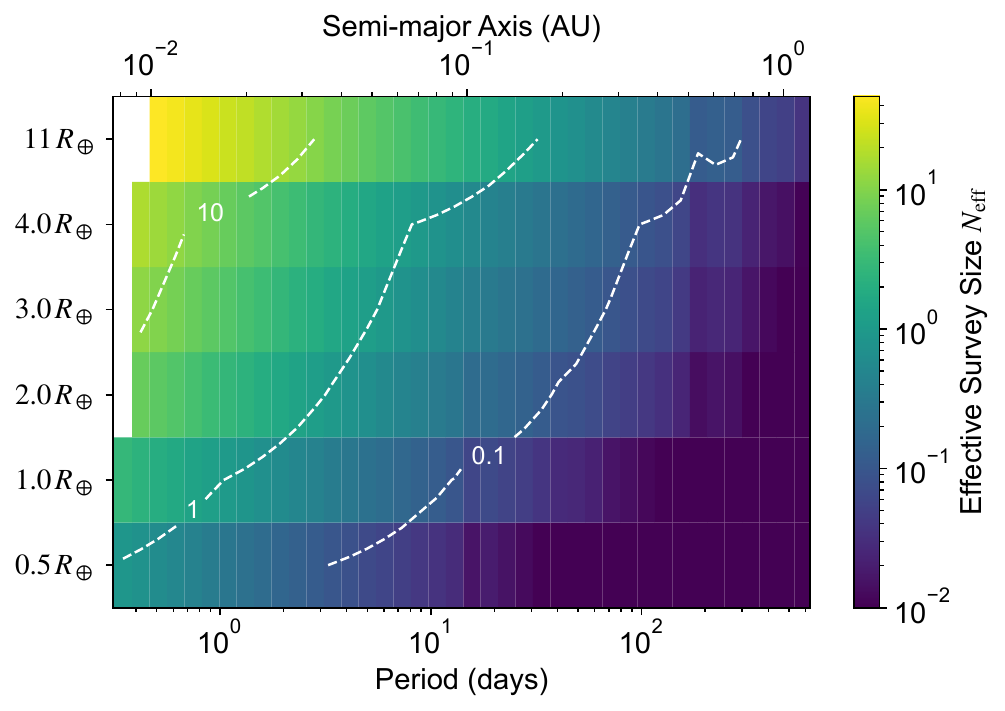}
    \caption{Effective survey size $N_{\rm eff}$ heat map as a function of orbital period and planet size. The white dashed lines show detection contours where 0.1, 1, and 10 detections are expected over the first three high-cadence GBTDS seasons. The blank cells on the upper left are excluded by the Roche limit.}
    \label{fig:neff_period_resolved_heatmap}
\end{figure}

While the period-averaged $\overline{N}_\mathrm{eff}$ predicts $\approx8$ detections of Jupiter-sized WD exoplanets by Roman GBTDS, a period-resolved $N_{\rm eff}$ simulation yields a more optimistic result (Figure \ref{fig:neff_period_resolved_heatmap}), because the higher $N_{\rm eff}$ contributions by the more easily detected close-in planets are not diluted by the longer period planets. The effective survey size of Jupiter-sized planets peaks at $N_{\rm eff}\approx47$ at an orbital period of $\approx0.5$ day, nearly six times the period-averaged value, corresponding to a required occurrence rate of $\eta_1\approx2.1\%$. At the $\approx1.4$-day orbit of WD~1856+534\,b \citep{vanderburg_giant_2020}, $N_{\rm eff}\approx20$ for Jupiter-sized planets ($\eta_1\approx4.9\%$). The peak effective survey size decreases toward smaller planets, to $N_{\rm eff}\approx16$, $11$, and $6.6$ ($\eta_1\approx6.1\%$, $8.9\%$, and $15.1\%$) for Neptune-, sub-Neptune-, and super-Earth-sized planets, respectively. At an orbital period of 500 min, $N_{\rm eff}\approx2.7$ ($\eta_1\approx36.6\%$) for Earth-sized planets, much higher than the period-averaged $\overline{N}_\mathrm{eff}$ value in Table \ref{tab:transit_yield} that is below unity. Even at their peak, Mars-sized planets remain below one expected detection, with $N_{\rm eff}\approx0.86$ ($\eta_1\approx116\%$).

The effective survey size $N_{\rm eff}$ as a function of period has a characteristic shape (Figure \ref{fig:expected_det_vs_eta_WD}b). Detectability rises steeply toward short periods, peaking at the shortest period allowed by the Roche limit. It falls monotonically toward long periods, as both the geometric transit probability (Section \ref{subsec:yield_framework}) and the number of observed transits decrease. On average, the peak $N_{\rm eff}$ at short periods is $\sim6\times$ greater than the period-averaged $\overline{N}_\mathrm{eff}$, implying that if short-period WD planets are indeed more common as opposed to the log-uniform period prior assumed earlier, Roman GBTDS will find $\sim47$ Jupiter-sized planets rather than $\approx8$.

This dramatic $\sim6\times$ difference demonstrates that the survey's sensitivity is dominated by the underlying period distribution of WD exoplanets, which is currently unconstrained. There are reasons to expect more detections of short-period giant planets similar to WD 1856+534 b, the only confirmed transiting WD exoplanet on a $\approx1.4$-day orbit \citep{vanderburg_giant_2020}. By modeling high-eccentricity migration driven by the Lidov-Kozai mechanism, \cite{munoz_kozai_2020} and \cite{oconnor_enhanced_2020} showed that a giant planet with an initial separation of a few to tens of AU around a main-sequence star can migrate onto a close-in orbit like that of WD 1856+534 b after the host evolves into a WD. \cite{munoz_kozai_2020} predicted that the LSST will find $\sim100$ close-in WD planets that have undergone this type of migration. However, WD planets can also be found on widely separated orbits. Population synthesis predicts that most surviving WD companions occupy wide orbits with a median separation of 11 AU \citep{mauch-soriano_predicted_2026}. Such widely separated planets will not transit more than once within the first three high-cadence seasons, even if their orbits are fortuitously aligned to be edge-on, and are therefore not detectable by Roman until the final three seasons offer us a longer baseline. While no confident constraints can be placed on the underlying period distribution of WD planets, our $N_\mathrm{eff}(R_p, P)$ grid can be folded with any such period distribution once more detections become available in the future.


\subsubsection{Host Brightness Dependence}
\label{subsubsec:brightness_dependence}

\begin{figure*}[t]
    \centering
    \includegraphics[width=\textwidth]{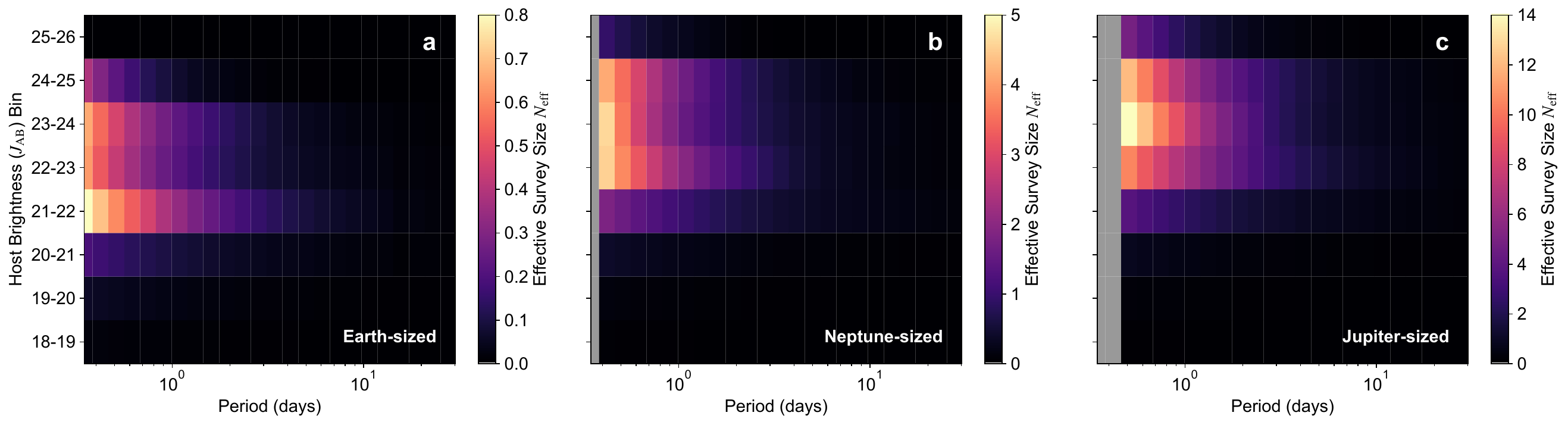}
    \caption{Effective survey size $N_{\rm eff}$ as a function of orbital period and WD host brightness $J_{\rm AB}$, for (a) Earth-sized ($1\,R_\oplus$), (b) Neptune-sized ($4\,R_\oplus$), and (c) Jupiter-sized ($11\,R_\oplus$) planets. The three panels use independent color scales. Gray cells at short period in (b) and (c) are excluded by the Roche limit.}
    \label{fig:neff_vs_host_brightness}
\end{figure*}

Here, we discuss how the effective survey size $N_{\rm eff}$ depends on host brightness, $J_{\rm AB}$. In principle, brighter WDs are the better targets due to better photometric precision and less severe source blending (Figure \ref{fig:source_blending}b). The GBTDS field WD population, however, is dominated by faint objects with a median brightness of $J_\mathrm{AB}\approx25.4$, and only $\approx2\%$ of hosts are brighter than $J_\mathrm{AB}=23$ (Figure \ref{fig:besancon_wd_population_histogram}). Therefore, the final yield per brightness bin depends on the balance between the per-WD detectability, which favors brighter hosts, and the number of available hosts, which rises steeply toward the faint end. In Figure \ref{fig:neff_vs_host_brightness}, we show the $N_\mathrm{eff}$ heatmap as a function of period and host brightness to reveal the optimal host brightness for each planet size.

For Earth-sized planets (Figure \ref{fig:neff_vs_host_brightness}a), the balance favors moderately bright hosts in the $J_\mathrm{AB}=21$--$22$ bin ($\approx33\%$ of the Earth yield) even though it holds only $79$ WDs. The faint WDs with $J_{\rm AB} \geq 24$, despite their vast numbers ($15{,}077$ out of $16{,}340$ in-field WDs), contribute only $\approx9\%$ to the overall $N_{\rm eff}$. Detections of terrestrial planets will therefore be confined to the bright targets.

Neptune- (Figure \ref{fig:neff_vs_host_brightness}b) and Jupiter-sized (Figure \ref{fig:neff_vs_host_brightness}c) planets, however, display the opposite pattern, with $N_{\rm eff}$ peaking in the relatively dim $J_{\rm AB}=23$--24 bin, implying that the number effect becomes more important for larger planets. The deep -- sometimes total -- transits of these larger planets remain detectable despite a moderate level of source blending. Nevertheless, if the host is too faint ($J_{\rm AB}=25$--26), detection is still difficult for the large planets due to severe blending ($D\sim10^{-3}$). In all cases, the brightest WDs ($J_\mathrm{AB}$ in the 18--21 bins) contribute modestly despite being individually the most favorable, because such bright WDs are intrinsically rare. Only $24$ of the $16{,}340$ in-field WDs are predicted to be this bright by our Besan\c{c}on population synthesis. Gaia observations also confirm the rarity of such WDs in the Galactic bulge. To date, only two WDs have been confirmed in the GBTDS fields according to the MWDD (Figure \ref{fig:besancon_gaia_in_GBTDS_fields}), and both have $G$ magnitudes $\sim19$.

\subsubsection{Eccentricity Dependence}
\label{subsubsec:ecc_dependence}

Close-in WD planets like WD 1856+534\,b could not have survived the giant phase at their present locations, so they must have been delivered inward from larger orbital separations by high-eccentricity migration. During such migration, a planet is scattered onto an eccentric orbit with a small pericenter and is then tidally circularized \citep{veras_tidal_2019, munoz_kozai_2020, oconnor_enhanced_2020}. Therefore, Roman may observe a planet still undergoing tidal circularization and retaining a nonzero eccentricity. We can use this eccentricity to measure how far the migration and circularization process has progressed relative to the WD cooling age \citep{veras_tidal_2019, li_can_2025_paper1}. Because the eccentricity distribution of WD exoplanets is unconstrained, the $\overline{N}_\mathrm{eff}$ values reported in Section~\ref{subsec:results_yield} assume $e=0$. Here, we relax that assumption to account for still-migrating planets, report the dependence of $\overline{N}_\mathrm{eff}$ on $e$, and assess the detectability of eccentric WD planets relative to circular ones.

\begin{figure}[t]
    \centering
    \includegraphics[width=\columnwidth]{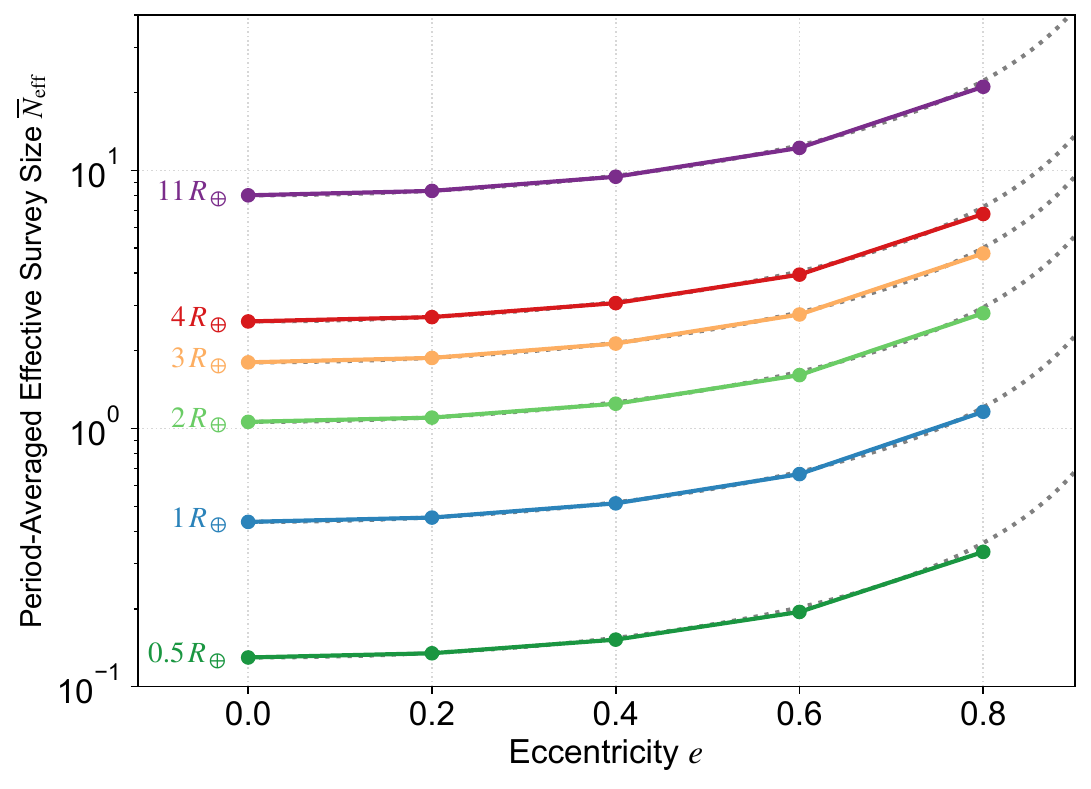}
    \caption{Period-averaged effective survey size $\overline{N}_\mathrm{eff}$ as a function of eccentricity $e$, color coded by planet size. Gray dotted curves show how the circular ($e=0$) values scale with the geometric transit probability, $P_{\rm geom} \propto 1/(1-e^2)$. The close agreement between colored and gray dotted curves confirms that the eccentricity dependence of $\overline{N}_\mathrm{eff}$ is dominated by the transit probability.}
    \label{fig:neff_vs_ecc}
\end{figure}

A nonzero eccentricity increases the effective survey size by a factor that is nearly the same for every planet size (Figure~\ref{fig:neff_vs_ecc}). For Jupiter-sized planets, the period-averaged $\overline{N}_\mathrm{eff}$ rises from $8.01$ on a circular orbit to $8.33$, $9.45$, $12.22$, and $21.09$ at $e = 0.2$, $0.4$, $0.6$, and $0.8$, respectively, a $\approx2.6\times$ increase at the highest eccentricity modeled. The same enhancement factor applies across the grid: at $e = 0.8$, every planet size gains a factor of $\approx2.6$ (Neptune-sized planets from $2.60$ to $6.78$, Earth-sized planets from $0.43$ to $1.16$), with smaller gains of $\approx1.04\times$, $1.18\times$, and $1.52\times$ at $e = 0.2$, $0.4$, and $0.6$, respectively. For Earth-sized planets, the highest eccentricity is enough to lift the $\overline{N}_\mathrm{eff}$ yield above one expected detection over the first three GBTDS seasons. This uniform enhancement across planet size is due to the increase in geometric transit probability, which has a factor of $1/(1-e^2)$ (see Section~\ref{subsec:yield_framework}). 
Because $\overline{N}_\mathrm{eff}$ rises monotonically with eccentricity, the values reported in Table \ref{tab:transit_yield} assuming $e=0$ are conservative lower bounds. If the underlying eccentricity distribution of WD planets is not concentrated at $e=0$, Roman GBTDS will be able to detect more transiting WD exoplanets than predicted in Section \ref{subsec:results_yield}.

\subsubsection{Favorable White Dwarf Properties}
\label{subsubsec:favorable_wds}

\begin{figure*}[t]
    \centering
    \includegraphics[width=\textwidth]{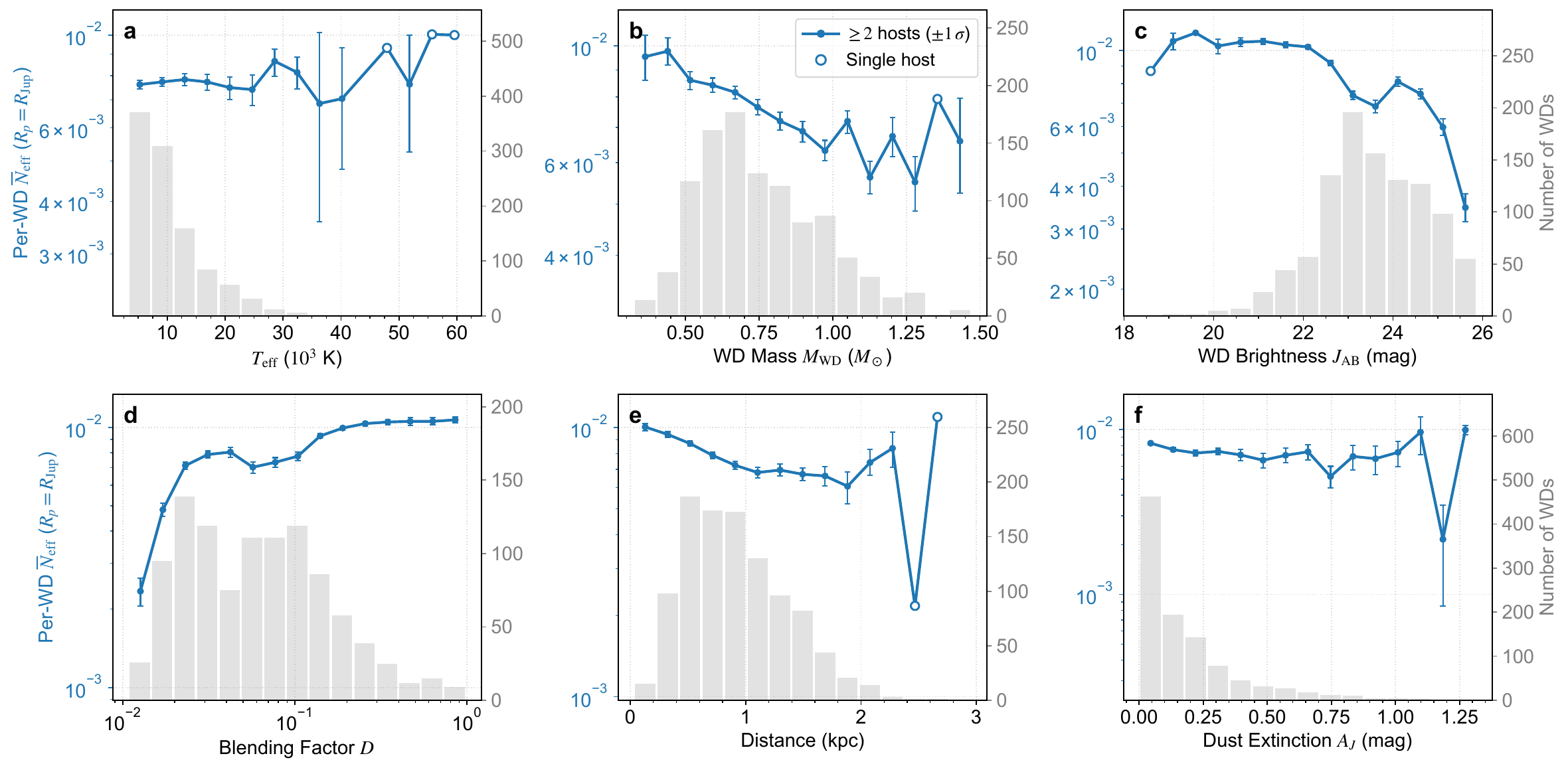}
    \caption{Per-WD $\overline{N}_\mathrm{eff}$ for Jupiter-sized planets (purple, left axis) as a function of six host properties: (a) effective temperature, (b) mass, (c) apparent brightness, (d) blending factor, (e) distance, and (f) dust extinction. Filled circles show the mean $\overline{N}_\mathrm{eff}$ over WDs in each bin, with $\pm1\sigma$ error bars. Open circles mark single-host bins, for which no error estimate is available. Gray histograms (right axis) show the number of detectable hosts (WDs with nonzero per-WD $\overline{N}_\mathrm{eff}$) per bin.}
    \label{fig:favorable_wd_properties}
\end{figure*}

Exoplanet surveys typically yield far more potential host stars than can be followed up individually, so it is essential to prioritize the most promising targets. TESS, for example, prioritized bright, cool dwarfs to maximize the yield of small transiting planets \citep{stassun_tess_2018}. Although Roman GBTDS will observe all in-field stars, resources for follow-up analysis are limited, so here we identify the properties of the most favorable WD hosts for a transiting planet search. We report the per-WD $\overline{N}_\mathrm{eff}$ as a function of six host properties (Figure~\ref{fig:favorable_wd_properties}): effective temperature ($T_{\rm eff}$), WD mass ($M_{\rm WD}$), WD brightness ($J_{\rm AB}$), blending factor ($D$), distance, and dust extinction ($A_J$). These per-WD $\overline{N}_\mathrm{eff}$ values are computed for Jupiter-sized planets, but the trends apply to all planet sizes.

High effective temperature is a favorable characteristic for WD hosts, as seen in the upward trend in Figure \ref{fig:favorable_wd_properties}a. A hotter WD is intrinsically more luminous, which raises the photon-limited SNR and hence the detection probability, so the mean per-WD $\overline{N}_\mathrm{eff}$ increases toward higher $T_\mathrm{eff}$. However, hot WDs are intrinsically rare. The synthetic WD population is overwhelmingly cool (see also Figure \ref{fig:besancon_wd_population_histogram}), so the high $T_{\rm eff}$ bins contain only a small number of WDs, leading to large per-WD $\overline{N}_\mathrm{eff}$ error bars. This tension will recur across many other host properties examined here: the WDs that individually offer the highest detectability are usually outliers in the population, so they serve as excellent priority targets but contribute little to the total yield. 

Lower mass WDs are better hosts (Figure~\ref{fig:favorable_wd_properties}b). Because WD radius increases as mass decreases, a lower $M_\mathrm{WD}$ raises the geometric transit probability and hence $\overline{N}_\mathrm{eff}$. In-field WD population peaks near $\sim0.6\,M_\odot$, at the low-mass end of the distribution, so the more detectable hosts are also among the more common ones.

Brighter WDs are preferred hosts, with higher per-WD $\overline{N}_\mathrm{eff}$ toward brighter $J_\mathrm{AB}$ (Figure~\ref{fig:favorable_wd_properties}c). The detectable hosts with nonzero per-WD $\overline{N}_\mathrm{eff}$ shown here peak near $J_\mathrm{AB}\approx23$, far brighter than the full Besan\c{c}on population, which peaks beyond $J_\mathrm{AB}\approx25$ (Figure~\ref{fig:besancon_wd_population_histogram}), because the faintest WDs do not contribute much to the overall detectability (see also Figure \ref{fig:neff_vs_host_brightness}).

WD hosts that suffer less from source blending are preferred (Figure~\ref{fig:favorable_wd_properties}d). The detectable WD host population is concentrated at $D\lesssim0.1$, so a typical host retains only a small fraction of its flux after blending, and the bins near $D\approx1$ that carry the highest per-WD $\overline{N}_\mathrm{eff}$ contain far fewer hosts than the severely blended bins, reflecting the same tension between favorable host properties and the number of available targets.

Nearby WD hosts are preferred, with per-WD $\overline{N}_\mathrm{eff}$ falling as distance increases (Figure~\ref{fig:favorable_wd_properties}e), because more distant WDs appear fainter, from both their greater distance and the larger dust column (see Figure \ref{fig:dust_extinction}c). The detectable hosts are concentrated within $\sim2$\, kpc with a peak just below 1 kpc. In contrast, the full Besan\c{c}on synthetic population extends beyond $\sim4$\, kpc (Figure~\ref{fig:besancon_wd_population_histogram}), so this $\sim2$\, kpc limit marks the depth to which Roman GBTDS can probe for transiting WD planets.

Dust extinction $A_J$ is included as a diagnostic rather than an independent parameter, because it is already folded into each WD's apparent magnitude $J_\mathrm{AB}$ from the Besan\c{c}on simulation. Per-WD $\overline{N}_\mathrm{eff}$ shows a weak decline toward higher $A_J$ (Figure~\ref{fig:favorable_wd_properties}f), though this trend is shallow and the uncertainties increase toward higher $A_J$, where such severely dust-extincted detectable hosts are rare. The number of detectable WD hosts peaks near $A_J\sim0$ and falls off quickly, well below the typical extinction of the GBTDS bulge fields, which peaks near $A_J\sim1$ mag (and near $A_J\sim4.5$ mag in the more heavily obscured Galactic center field).

Varying the thickness of the WD hydrogen envelope has a negligible effect on the predicted transit yield. At fixed mass and $T_\mathrm{eff}$, a thick-envelope WD is puffier than a thin-envelope WD. To assess the difference, we simulate a subgrid with a thin H envelope (Table~\ref{tab:grid}), while holding all other grid parameters fixed. A thin H envelope lowers $\overline{N}_\mathrm{eff}$, but by less than $3\%$ for all planet sizes, and by only $\approx0.3\%$ for Jupiter-sized and $\approx0.7\%$ for Neptune-sized planets that dominate the yield. Envelope thickness therefore has a negligible effect on the yield predictions.

In summary, the six host properties shown in Figure \ref{fig:favorable_wd_properties} reduce to two mechanisms that determine detectability: WD mass, which sets the radius and hence the geometric transit probability, and photon flux, which sets the SNR and hence the detection probability $P_{\rm det}$. The most favorable WDs are therefore low-mass (large radius), hot, bright, nearby, not severely blended, and not severely dust-extincted. However, except for WD mass, whose favorable low-mass end coincides with the peak of the population, WDs with favorable properties tend to be rare. The favorable hosts are therefore best understood as priority targets rather than as the source of the bulk transit detection yield, which will come from the far more numerous typical hosts: cool ($\lesssim10^4$ K), $\sim0.6\, M_\odot$, and faint.

\subsection{Constraints on the WD Exoplanet Occurrence Rate from a Null Detection}
\label{subsec:null_result}

Given that only one transiting WD exoplanet has been confirmed to date \citep{vanderburg_giant_2020}, and that a survey of over 1000 WDs by K2 yielded a null detection \citep{van_sluijs_occurrence_2018}, it is possible that such exoplanets are intrinsically rare and that Roman GBTDS will also yield a null detection. A null result is nonetheless scientifically informative, since it places an upper limit on the WD exoplanet occurrence rate ($\eta_\mathrm{WD}$) through the statistical formalism discussed in Section~\ref{subsubsec:null_method}.

A Roman GBTDS null detection would most tightly constrain the occurrence rate of giant planets. Averaged over period, a null detection would limit the occurrence of Jupiter-sized planets to $\eta_\mathrm{WD} < 28\%$ and of Neptune-sized planets to $< 57\%$ (all $\eta_\mathrm{WD}$ limits are reported at $95\%$ confidence). The limit weakens steadily toward smaller planets, whose shallower transits are intrinsically harder to detect, so that a null result can barely distinguish a true absence of planets from a failure to detect planets that exist. For sub-Neptune-, super-Earth-, and Earth-sized planets, a null detection would constrain the occurrence rate to $< 66\%$, $< 77\%$, and $< 88\%$, respectively, and for Mars-sized planets to $< 93\%$, which is effectively unconstrained. Occurrence rate upper limits provided by Roman GBTDS will be the first such constraints on WD planets in the Galactic bulge and center regions, complementing existing limits derived from nearby WD populations (see Section~\ref{subsec:survey_comparison} for further discussion).

\begin{figure}[t]
    \centering
    \includegraphics[width=\columnwidth]{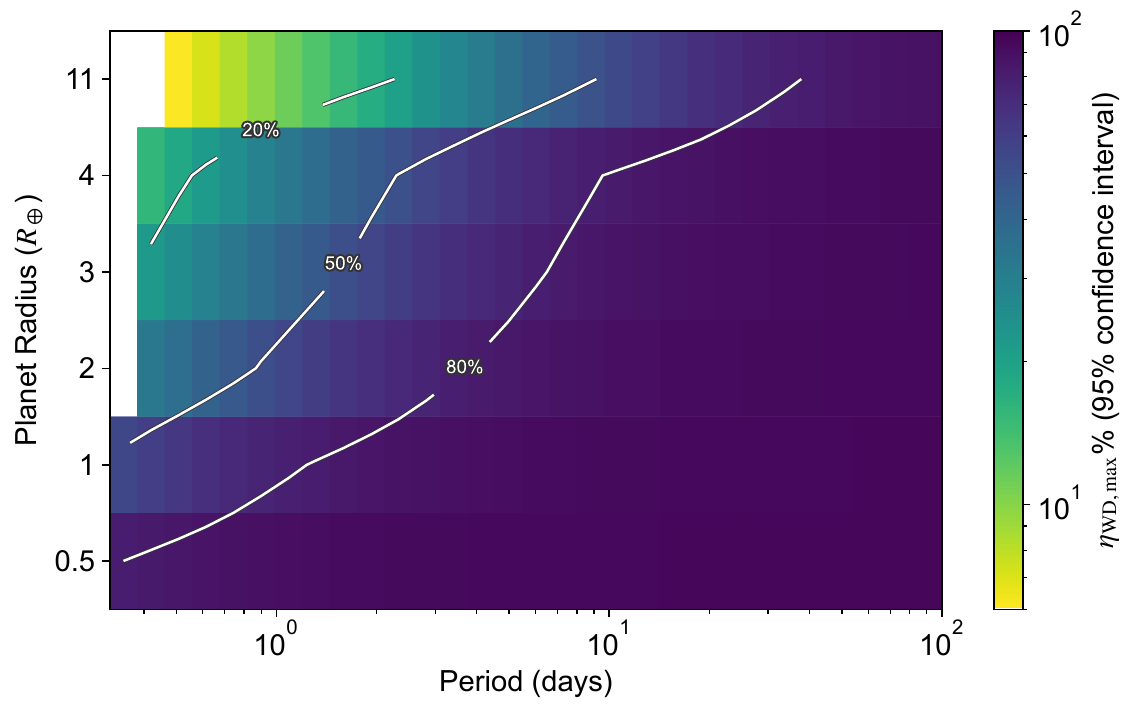}
    \caption{WD exoplanet occurrence-rate upper limit $\eta_\mathrm{WD, max}$ (95\% confidence interval) from a Roman GBTDS null detection, in the radius-period plane. White contours mark $\eta_\mathrm{WD,max} = 20\%$, $50\%$, and $80\%$. Blank regions are unphysical and excluded by the Roche limit.}
    \label{fig:null_result_limit_on_eta}
\end{figure}

When period resolution is available, WD planet occurrence rate can be constrained more tightly than the period-averaged limits, because the survey is much more sensitive at shorter periods. In Figure~\ref{fig:null_result_limit_on_eta}, we present a heatmap of the upper limit $\eta_\mathrm{WD, max}$ in the radius-period plane, without any assumption about the underlying period distribution. The limit is tightest for large planets on short-period orbits, where the effective survey size is highest. It relaxes toward longer periods and smaller planet sizes, tracing the same sensitivity structure as the detection yield (Section~\ref{subsubsec:period_dependence}). Two regimes are particularly interesting. For close-in Jupiter-sized planets just outside the Roche limit ($P \approx 0.5$\,d), resembling WD~1856+534\,b, a null detection would constrain the occurrence rate to $< 6\%$. For Earth-sized planets at the shortest sampled period, a rough proxy for the WD habitable zone, the limit is $< 55\%$.

Source blending, rather than the size of the WD sample, is the dominant limitation on these $\eta_\mathrm{WD}$ constraints. 
Without blending, the same null detection would constrain Jupiter-sized planets to $\eta_\mathrm{WD} < 1.7\%$, roughly $16$ times tighter than the blended limit of $< 28\%$ (assuming a log-uniform period distribution).

\subsection{Injection-Recovery Test Validates Grid Results}
\label{subsec:injection_results}

\begin{figure}[t]
    \centering
    \includegraphics[width=\columnwidth]{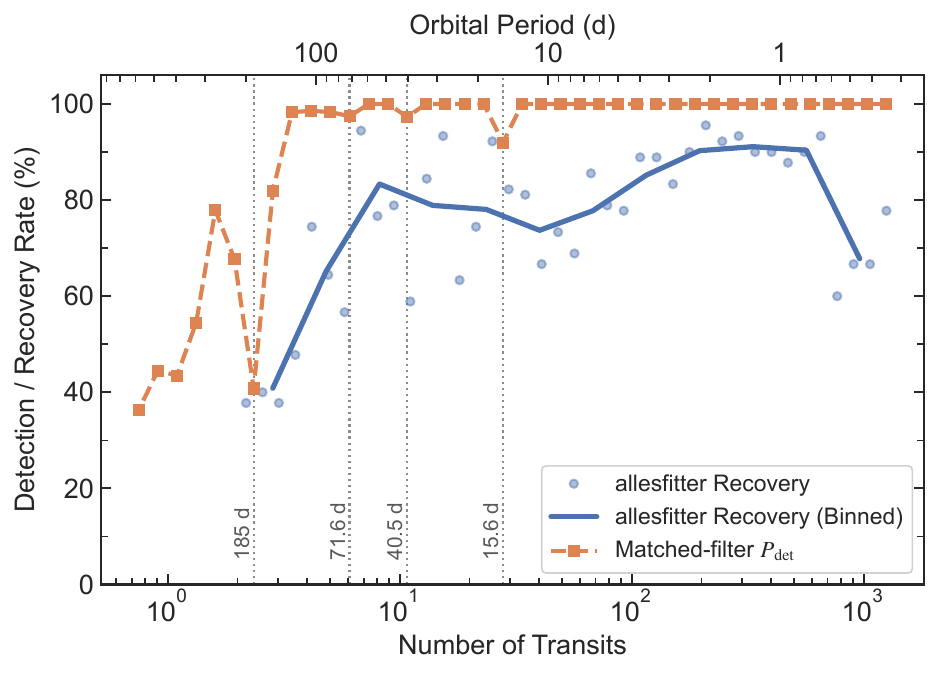}
    \caption{Recovery fraction of injection-recovery test (\texttt{allesfitter}) and matched-filter detection probability $P_\mathrm{det}$ as functions of the number of transits ($N_\mathrm{transit}$) within the survey baseline. The top axis shows the corresponding orbital period. Blue points show the recovery fraction per injected cell, and the blue line its binned trend. The dashed orange line is the matched-filter $P_\mathrm{det}$. Both detection and recovery fractions decline toward small $N_\mathrm{transit}$, but the blind search requires $\sim3\times$ as many transits to reach the same completeness, highlighting the penalty for searching without a known ephemeris. Note that at the shortest periods, the transit is readily detected, but its period is frequently aliased to a harmonic, so the recovery fraction drops. The vertical dotted lines show where detectability decreases because of aliasing. The 71.6 d period is close to the 70.5-day season length, and the 185 d period roughly coincides with the season interval. Dips at the 15.6- and 40.5-day periods occur because transits there happen at nearly the same phase as the exposure.}
    \label{fig:injection_recovery}
\end{figure}

To validate that our matched-filter SNR grid results (Section~\ref{subsec:grid}) survive a realistic blind search with no prior knowledge of the transit time and depth, we perform an injection-recovery test with \texttt{allesfitter} (Section~\ref{subsec:injection}). The blind search reproduces the detectability structure of the SNR grid. Across all $3600$ injections, we recovered $76\%$ of the transits. The detection and recovery fractions of both methods follow the same trend as a function of the number of transits ($N_\mathrm{transit}$) within the $436$-day survey baseline (Figure~\ref{fig:injection_recovery}), losing sensitivity as $N_\mathrm{transit}$ drops as $P$ increases.

The \texttt{allesfitter} injection-recovery test requires $\sim3\times$ more transits than the matched-filter grid to reach the same recovery fraction, as expected for a blind search that must solve for unknown transit ephemeris. This penalty is more severe at longer periods when transits are rare. Blind completeness reaches $90\%$ by $N_\mathrm{transit} \approx 190$ ($P \approx 2.3$\,d) and falls below $50\%$ for $P \gtrsim 122$\,d, dropping to $\sim38\%$ at the longest simulated period. Because the transit yield is weighted toward short periods by the geometric transit probability, our yield prediction is robust to the transition from the idealized grid to a realistic blind search. At the shortest periods, the recovery fraction declines somewhat unexpectedly (Figure~\ref{fig:injection_recovery}), but this reflects a failure to constrain the precise period rather than a failure to detect the transit, because the recovered periods at these very close-in orbits are frequently aliased to a harmonic.

Our injection-recovery test uses a single, bright ($J_{\rm AB}=20$) host and varies only orbital period and planet radius (Section~\ref{subsec:injection}). Thus, it isolates the ephemeris-search penalty rather than a photon-noise limit. Fainter hosts would lower both the detection and recovery fraction, but we expect them to follow the same trend as shown in Figure \ref{fig:injection_recovery}. A full characterization of pipeline completeness as a function of host brightness is left to future work.

\subsection{Secondary Eclipse Does Not Contribute to Transit Yields}
\label{subsec:eclipse_null}

Secondary eclipses do not contribute to the overall detection probability, according to our simulation using an edge-on ($b = 0$) eclipse SNR grid that matches the transit base grid in the other axes (Table~\ref{tab:grid}). Eclipse detectability is confined to a very narrow corner of short-period, large-planet, bright, hot-WD systems, with a maximum SNR of $\approx 33$ in the most favorable cell, and only $\sim$0.01\% of grid cells reach $P_\mathrm{det} \ge 0.5$. Across the shared base grid of $1{,}872{,}000$ cells, the eclipse-to-transit SNR ratio is $\le 0.78\%$, so combining the two signals 
boosts the transit SNR by $\le 0.003\%$ and does not rescue any marginal transit detections. Therefore, we ignore the secondary-eclipse channel entirely and report only the detectability of primary transits.

\subsection{Transit Detection Yield Framework}
\label{subsec:yield_framework}

Here, we explain in detail how we calculate the three quantities reported above: the effective survey size $N_\mathrm{eff}$, the occurrence rate ${\eta}_1$ required for Roman GBTDS to expect one detection (Section~\ref{subsubsec:neff_method}), and the upper limit on $\eta_{\rm WD}$ that a null result by GBTDS would impose (Section~\ref{subsubsec:null_method}).

\subsubsection{Effective Survey Size and Required Occurrence Rate}
\label{subsubsec:neff_method}

To obtain the effective survey size, we map each of the $16{,}340$ in-field WDs from the Besan\c{c}on population synthesis to its nearest grid bin in $(J_\mathrm{AB}, T_\mathrm{eff}, M_\mathrm{WD})$, and define $N_{\rm eff}$ at a given planet radius and orbital period as a sum
\begin{equation} \label{eq:neff}
    N_\mathrm{eff}(R_p, P) = \sum_i P_{\mathrm{geom},i}(P)\,P_{\mathrm{det},i}(P),
\end{equation}
following a previous work \citep{van_sluijs_occurrence_2018}, with the only difference being that we define transit probability $P_{\mathrm{geom},i}$ and detection probability $P_{\mathrm{det},i}$ for each individual WD $i$. This equivalence allows us to adopt the same null-result formalism from \cite{van_sluijs_occurrence_2018} in Section~\ref{subsubsec:null_method}. The expected number of detections for an underlying WD exoplanet occurrence rate $\eta_\mathrm{WD}$ is $N_\mathrm{det} = \eta_\mathrm{WD}\, N_\mathrm{eff}$, so a survey of $N_\mathrm{eff}$ effective WDs expects one detection if the intrinsic occurrence rate of WD exoplanets is $\eta_1(R_p, P) = 1/N_\mathrm{eff}$.

Averaging over a uniform distribution of argument of periastron $\omega$, the geometric transit probability of a WD exoplanet is \citep{barnes_effects_2007, burke_impact_2008}
\begin{equation} \label{eq:pgeom}
    P_\mathrm{geom} = \frac{R_\mathrm{WD} + R_p}{a}\cdot\frac{1}{1-e^2},
\end{equation}
where we derive $R_\mathrm{WD}$ from the \cite{bedard_spectral_2020} cooling tracks at each WD's $(T_\mathrm{eff}, M_\mathrm{WD})$ and the semi-major axis $a$ from Kepler's third law. Orbital eccentricity $e$ remains a free parameter.

The detection probability $P_{\mathrm{det}, i}$ is evaluated at each WD's grid point, which is defined as the fraction of MC trials recovered with $\mathrm{SNR} \ge 7$ (Section~\ref{subsec:grid}).

To quantify the survey sensitivity with a single number per planet size, we average $N_\mathrm{eff}(R_p, P)$ over a log-uniform orbital period prior,
\begin{equation} \label{eq:neff_int}
    \overline{N}_\mathrm{eff}(R_p) = \frac{\sum_j N_\mathrm{eff}(R_p, P_j)\,\Delta\log P_j}{\log P_\mathrm{range}},
\end{equation}
where the overbar denotes the average over the log-uniform period range and $\log P_\mathrm{range}$ is the entire range (from 500 min to 581 d, see Table \ref{tab:grid}) that we consider. Marginalizing over period requires a period prior, so we adopt a log-uniform grid, but note that this is an assumption because the underlying period distribution of WD exoplanets is unknown. Similarly, marginalizing over eccentricity would require an eccentricity prior, which is unconstrained for WD exoplanets, so we instead fix $e = 0$ when computing $\overline{N}_\mathrm{eff}$. 
We report $\overline{N}_\mathrm{eff}$ in Section \ref{subsec:results_yield} as the overall conservative estimate, but also present the period-resolved $N_\mathrm{eff}(R_p, P)$, which retains the full short-period sensitivity, in Section \ref{subsubsec:period_dependence}. The corresponding period-averaged occurrence rate for one expected detection is $\overline{\eta}_1 = 1/\overline{N}_\mathrm{eff}$.


\subsubsection{Calculating the Upper Limit on WD Exoplanet Occurrence Rate from a Null Detection}
\label{subsubsec:null_method}

In Section \ref{subsec:null_result}, we report that Roman GBTDS can constrain the underlying WD exoplanet occurrence rate $\eta_{\rm WD}$, even if it detects zero transiting WD planets. Here, we derive the expression for this upper limit, following the binomial null detection formalism that \cite{van_sluijs_occurrence_2018} applied to the K2 null result.

For a null detection from an effective survey size of $N_\mathrm{eff}$, the maximum occurrence rate consistent with the nondetection at confidence level $C$ is
\begin{equation} \label{eq:fmax}
    \eta_\mathrm{WD,max} = 1 - (1 - C)^{1/(N_\mathrm{eff} + 1)}.
\end{equation}
We quote $95\%$ confidence limits ($C = 0.95$) throughout. We note a nuance that $\eta_\mathrm{WD,max}$ is mathematically different from the $\overline{\eta}_1 = 1/\overline{N}_\mathrm{eff}$ rate reported above. By construction, $\eta_\mathrm{WD, max} < C \le 1$, whereas $\overline{\eta}_1$ can exceed unity when $\overline{N}_\mathrm{eff} < 1$, implying that multiple planets in a system are required for one detection. While not equivalent, the two quantities scale with $\overline{N}_\mathrm{eff}$ in the same way and differ by a factor set by the confidence level $C$, and are therefore roughly comparable.


\section{Non-Transiting Planet (Phase Curve) Detection Yields}
\label{sec:results_nontransit}

Roman can also detect WD exoplanets that do not transit, by probing their thermal phase curves. Any planet orbiting a WD re-emits part of the incident radiation as thermal infrared emission and reflects part of it, producing a periodic phase curve modulation that we model as the sum of the two components (Section~\ref{sec:nontransit_model}). This modulation is strongest for close-in planets around hot WDs. Unlike a transit, which requires precise orbital alignment, this periodic modulation is visible across nearly all orbital inclinations, so the phase curve channel can in principle access a much larger WD population. This approach has precedent for main-sequence hosts. \cite{millholland_supervised_2017} used photometric phase curve modulations to search for non-transiting planets in the Kepler sample, reporting sixty non-transiting hot Jupiter candidates.


Here, we compute the phase curve yield over the same $16{,}340$ in-field WDs used for the transit channel (Section~\ref{subsec:population}), so the two channels are directly comparable. We report the overall predicted yield in Section~\ref{subsec:results_yield_nontransit}, examine where in the system parameter space the yield concentrates in Section~\ref{subsec:pc_resolved}, and quantify the phase curve's contribution in addition to the transit channel in Section~\ref{subsec:complementarity}. We introduce the phase curve yield calculation method at the end in Section~\ref{subsec:pc_framework}.

\subsection{Phase Curve Detection Yield}
\label{subsec:results_yield_nontransit}

\begin{deluxetable}{lcccc}
\tabletypesize{\scriptsize}
\tablewidth{\columnwidth}
\tablecaption{Effective survey size marginalized over log-uniform period bins ($\overline{N}_\mathrm{eff}^\mathrm{PC}$) for non-transiting (phase curve) WD exoplanets in the GBTDS, by planet size, for the raw, Roche limit, and the final Roche plus source blending tiers. The final column gives $\overline{\eta}_1 = 1/\overline{N}_\mathrm{eff}^\mathrm{PC}$, the occurrence rate required for one expected detection. \label{tab:pc_yield}}
\tablehead{
\colhead{Planet size} & \colhead{$\overline{N}_\mathrm{eff}^\mathrm{PC}$ raw} & \colhead{$+$Roche} & \colhead{$+$blend} & \colhead{$\overline{\eta}_1$}
}
\startdata
Mars (0.5\,$R_\oplus$)        & $\sim$0  & $\sim$0  & $\sim$0  & --        \\
Earth (1\,$R_\oplus$)         & $\sim$0  & $\sim$0  & $\sim$0  & --        \\
Super-Earth (2\,$R_\oplus$)   & 0.0015   & $\sim$0  & $\sim$0  & --        \\
Sub-Neptune (3\,$R_\oplus$)   & 0.117    & 0.0237   & 0.0131   & $>$1000\% \\
Neptune (4\,$R_\oplus$)       & 0.544    & 0.183    & 0.0617   & $>$1000\% \\
Jupiter (11\,$R_\oplus$)      & 54.5     & 10.4     & 0.351    & 285\%     \\
\enddata
\end{deluxetable}

Phase curve observations by Roman GBTDS will provide at best a marginal detection of WD exoplanets, and the limited sensitivity is confined to a narrow parameter space of close-in giant planets. 

\begin{figure}[t]
    \centering
    \includegraphics[width=\columnwidth]{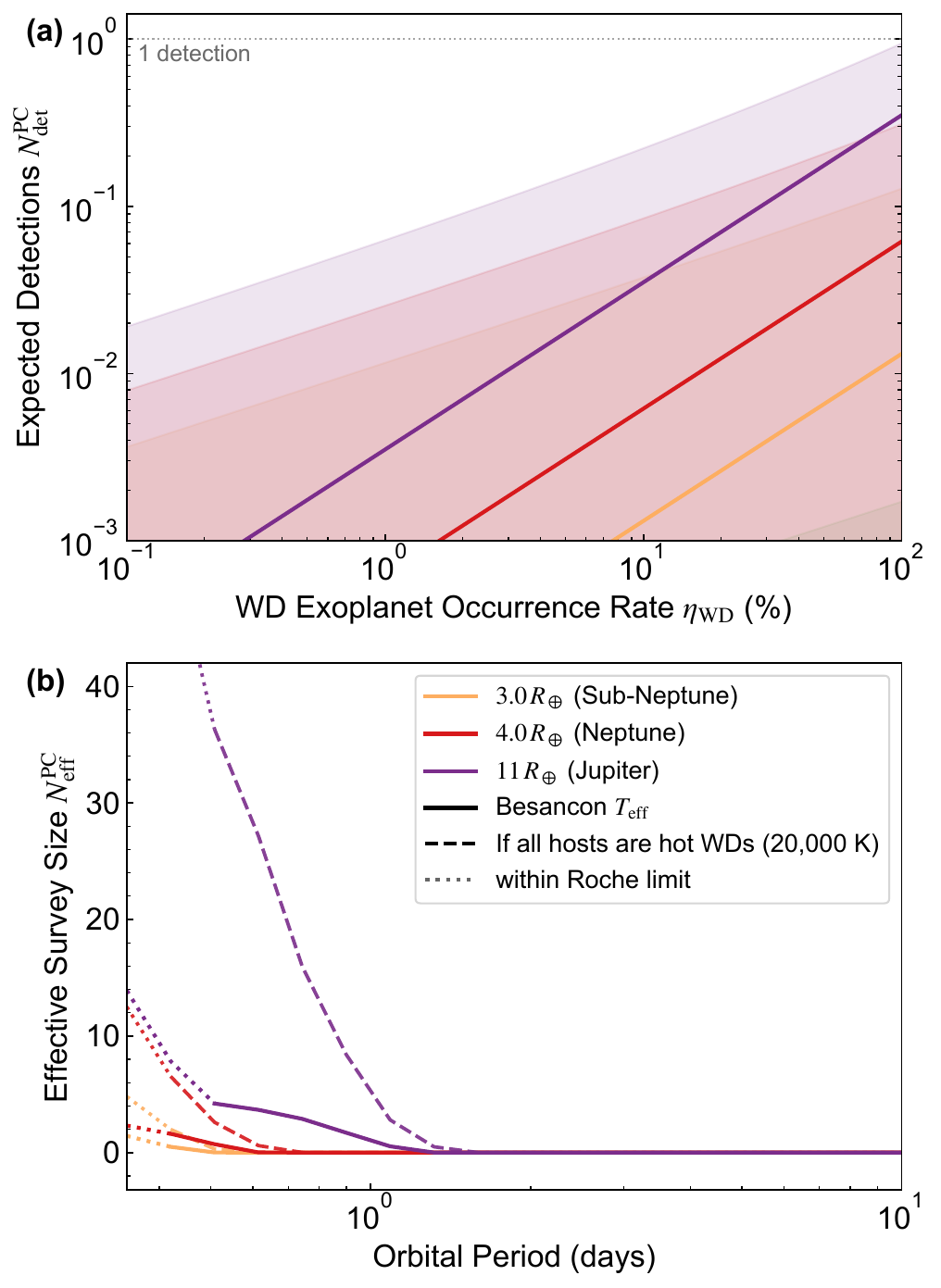}
    \caption{(a) Expected number of phase curve WD exoplanet detections by Roman GBTDS, $N_{\rm det}^{\rm PC}$, as a function of the underlying WD exoplanet occurrence rate $\eta_{\rm WD}$, color coded by planet size. 
    The shaded areas show $\pm1\sigma$ Poisson uncertainties. (b) Effective survey size $N_{\rm eff}^{\rm PC}$ as a function of planetary orbital period. $N_{\rm eff}^{\rm PC}$ peaks just outside the Roche distance and decreases towards longer orbital period. Solid curves use each WD's $T_\mathrm{eff}$ from the Besan\c{c}on population synthesis. Dashed curves show a hypothetical scenario in which every host is hot ($T_\mathrm{eff} = 20{,}000$\, K), illustrating the sensitivity of the phase curve yield to host temperature. Periods that are too short and within the Roche limit are demarcated as dotted curves.}
    \label{fig:yield_nontransiting}
\end{figure}

Overall, no planet size reaches one expected detection. Jupiter-sized ($11\,R_\oplus$) planets have the largest phase curve yield, $\overline{N}_\mathrm{eff}^\mathrm{PC}\approx0.35$, followed by Neptune-sized ($4\,R_\oplus$) and sub-Neptune-sized ($3\,R_\oplus$) planets with $\overline{N}_\mathrm{eff}^\mathrm{PC}\approx0.06$ and $\approx0.013$, respectively (Table~\ref{tab:pc_yield} and Figure \ref{fig:yield_nontransiting}a). Super-Earth-sized and smaller planets are effectively undetectable through their phase curves ($\overline{N}_\mathrm{eff}^\mathrm{PC}\approx0$). In terms of the underlying occurrence rate required for one expected detection over the first three high-cadence seasons, Jupiter-sized planets need $\overline{\eta}_1\approx285\%$ (nearly three giant planets per WD), while smaller planets require unphysical occurrence rates $>1000\%$ (Table~\ref{tab:pc_yield}).

Here, the phase curve effective survey size $\overline{N}_\mathrm{eff}^\mathrm{PC}$ is the expected number of detections if every one of the $16{,}340$ in-field WDs hosts exactly one planet of the given radius. Unlike a transit search, which probes only nearly edge-on systems, a phase curve is detectable across a wide range of inclinations, so we average each WD's detectability over an isotropic prior uniform in $\cos i$, and then over the log-uniform period prior.

Phase curve yields are substantially smaller than the transit yields (Section~\ref{subsec:results_yield}), because the signal strength severely limits the former. Even when assuming the most favorable system configuration, the two channels differ by a factor of $\approx50$. Consider the first period bin outside of the Roche limit for a Jupiter-sized planet ($P = 12.2$\,hr) around an optimistic host: a hot ($T_\mathrm{eff} = 20{,}000$\, K), low mass ($0.4\, M_\odot$), bright ($J_\mathrm{AB} = 18$) WD. The planet's dayside reaches $T_\mathrm{day}\approx1560$\, K, and the phase curve amplitude reaches $\approx2\%$ from peak to trough at an orbital inclination of $i = 80^\circ$. Assuming the inclination is $i = 90^\circ$ in the same system (perfectly edge-on), the planet would occult the WD completely, producing a $100\%$ dip, which is $\approx50\times$ deeper than the phase curve amplitude. 
Cooler, fainter, and more massive WD hosts yield weaker sinusoidal modulations, and source blending in the crowded bulge fields dilutes the remaining weak signal (Section~\ref{subsubsec:blending}), so $\approx2\%$ is only an optimistic upper limit for close-in giants orbiting those rare hot, bright, weakly blended WD hosts.

The Roche limit shapes the phase curve yield far more strongly than the transit yield, because the phase curve signal amplitude itself depends on the planetary dayside emission temperature, which scales as $T_\mathrm{day}\propto a^{-1/2}$ \citep{cowan_statistics_2011}. For Jupiter-sized planets, the Roche cut decreases $\overline{N}_\mathrm{eff}^\mathrm{PC}$ from $54.5$ to $10.4$, an $\approx81\%$ loss, whereas the same cut costs the transit channel only $\approx18\%$ (Table~\ref{tab:transit_yield}). Source blending then removes a further $\approx97\%$ ($10.4$ to $0.351$), comparable to the $\approx95\%$ it removes from the transit yield.

\subsection{Where the Phase Curve Yield Concentrates}
\label{subsec:pc_resolved}

The $\overline{N}_\mathrm{eff}^\mathrm{PC}$ values reported above are averaged over a log-uniform period prior and an isotropic orbital inclination prior, so they do not reveal where in the system parameter space the surviving phase curve detectability concentrates. Here, we resolve the yield along the same three axes as the transit channel to allow direct comparison: orbital period (Section~\ref{subsubsec:pc_period_dependence}), host brightness (Section~\ref{subsubsec:pc_brightness_dependence}), and eccentricity (Section~\ref{subsubsec:pc_ecc}).

\subsubsection{Orbital Period Dependence}
\label{subsubsec:pc_period_dependence}

The phase curve yield is confined to a narrow window of close-in orbits. The effective survey size of Jupiter-sized planets peaks at $N_\mathrm{eff}^\mathrm{PC}\approx4.2$ at an orbital period of $\approx0.5$\,d, immediately outside the Roche limit, then falls by more than three orders of magnitude over a factor of $\approx2.6$ in period, dropping below $N_\mathrm{eff}^\mathrm{PC}=10^{-3}$ by $P\approx1.3$\,d (Figure~\ref{fig:yield_nontransiting}b). The effective survey sizes of smaller planets peak lower and fall off sooner. Neptune-sized planets reach $N_\mathrm{eff}^\mathrm{PC}\approx1.6$ and sub-Neptune-sized planets $N_\mathrm{eff}^\mathrm{PC}\approx0.49$, both at $P\approx0.42$\,d, again just outside the Roche limit. Their detectability drops below $N_\mathrm{eff}^\mathrm{PC}=10^{-3}$ at $P\approx0.6$\,d and $\approx0.5$\,d, respectively.

In addition to planet size and orbital period, the host effective temperature is another important factor in the phase-curve yield. To illustrate this sensitivity, we recompute the yield for a hypothetical scenario in which every in-field WD is a hot $20{,}000$\, K host, holding its synthetic $J_\mathrm{AB}$ and $M_\mathrm{WD}$ fixed (Figure~\ref{fig:yield_nontransiting}b, dashed lines). The peak Jupiter-sized $N_\mathrm{eff}^\mathrm{PC}$ rises from $\approx4.2$ to $\approx36$, a factor of $\approx8.6$ increase. Neptune- and sub-Neptune-sized planets peak at $N_\mathrm{eff}^\mathrm{PC}\approx6.5$ and $\approx2.0$ rather than $\approx1.6$ and $\approx0.49$.

The hypothetical $20{,}000$\, K scenario is not as unphysical as it may appear, because the phase curve yield is already heavily concentrated in the hottest hosts. Among the $16{,}340$ GBTDS field WDs, $3{,}104$ are hotter than $16{,}000$\, K (roughly $19\%$ of the sample), but these rare hot WDs supply $\approx98\%$ of the period-averaged Jupiter-sized yield. Our base grid has an upper limit of $T_\mathrm{eff} = 20{,}000$\, K (Table~\ref{tab:grid}), so the $1{,}653$ in-field WDs hotter than this ($10.1\%$ of the sample, extending to $\approx75{,}000$\, K) are evaluated at the grid ceiling rather than at their true temperatures. These clipped hosts supply $\approx47\%$ of the period-averaged Jupiter-sized yield, and their true detectability is higher than what was calculated using the $20{,}000$\, K proxy. Two effects combined to make hot hosts favorable. A hotter WD irradiates its planet more strongly, raising the dayside temperature and hence the planet's emission in the \texttt{F146} band. At the same time, a hot WD's SED peaks at short wavelengths, so the host's own brightness in the band grows more slowly than the planetary flux as $T_\mathrm{eff}$ rises.

The prospect of finding exoplanets orbiting hot WDs goes beyond the GBTDS. Because a phase curve is a continuous modulation rather than a time-sensitive event, it does not require the 12.1-minute high cadence. The phase-curve channel may therefore be better suited to the High-Latitude Time-Domain Survey (HLTDS). We discuss this possibility in Section \ref{subsec:high_latitude}.

\begin{figure*}[t]
    \centering
    \includegraphics[width=\textwidth]{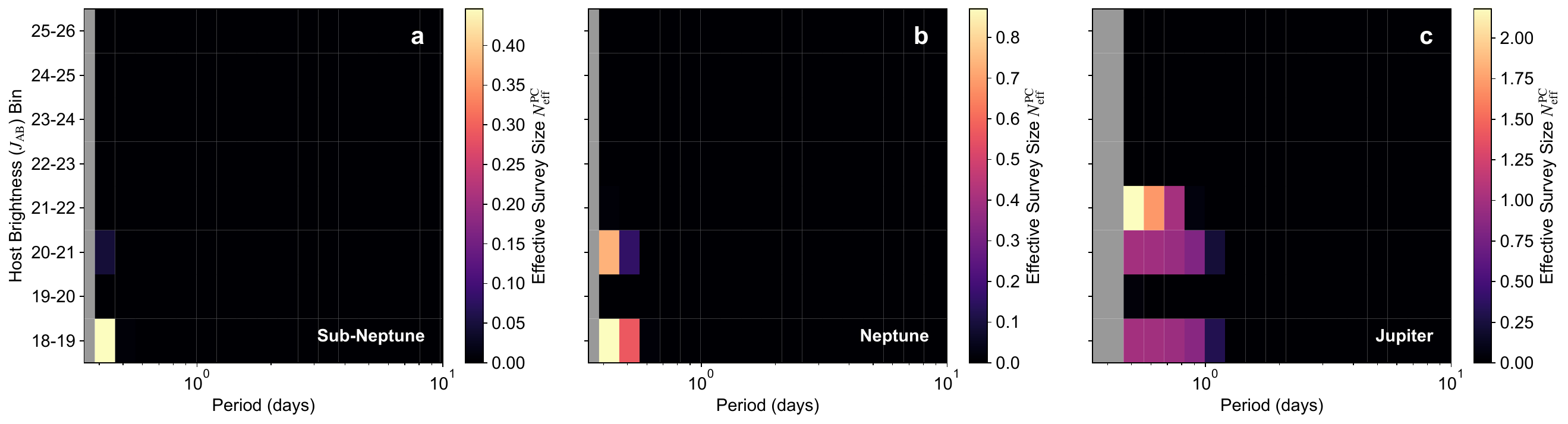}
    \caption{Effective survey size $N_{\rm eff}^{\rm PC}$ as a function of orbital period and WD host brightness $J_{\rm AB}$, for (a) sub-Neptune-sized ($3\,R_\oplus$), (b) Neptune-sized ($4\,R_\oplus$), and (c) Jupiter-sized ($11\,R_\oplus$) planets. The three panels use independent color scales. Gray cells at short period are excluded by the Roche limit. For all three sizes, the phase curve yield is confined to short-period orbits around bright hosts, while fainter WDs contribute negligibly.}
    \label{fig:neff_vs_host_brightness_pc}
\end{figure*}

\subsubsection{Host Brightness Dependence}
\label{subsubsec:pc_brightness_dependence}

Most phase curve detections from Roman GBTDS are expected from bright hosts. Nearly the entire surviving Jupiter-sized yield comes from three brightness bins ($J_\mathrm{AB} = 18$--19, 20--21, and 21--22), which contribute $\approx31\%$, $\approx30\%$, and $\approx38\%$ of the total yield, respectively. In contrast, everything fainter than $J_\mathrm{AB} = 22$ contributes $\approx0.05\%$ in total (Figure \ref{fig:neff_vs_host_brightness_pc}). This is the opposite of the transit channel, where the deep transits of Neptune- and Jupiter-sized planets remain detectable around fainter hosts and the yield peaks in the $J_\mathrm{AB} = 23$--24 bin (Section~\ref{subsubsec:brightness_dependence}). Because phase curve signals have small intrinsic amplitudes (Section~\ref{subsec:results_yield_nontransit}), only bright hosts offer the photometric precision a detection requires. 

As the planet sizes shrink, detectability becomes more concentrated in the brighter hosts despite their rarity. Approximately $62\%$ of the surviving yield for Neptune-sized planets is concentrated in the $J_\mathrm{AB} = 18$--19 bin, while the remaining $\approx38\%$ lies in the $J_\mathrm{AB} = 20$--21 bin (the 19--20 bin contributes nothing despite holding more hosts than the 18--19 bin, see Figure~\ref{fig:neff_vs_host_brightness_pc}b). Sub-Neptune-sized planets are even more concentrated, with $\approx91\%$ of the yield from the $J_\mathrm{AB} = 18$--19 bin and the remaining $\approx9\%$ from 20--21 (Figure~\ref{fig:neff_vs_host_brightness_pc}a). The bright end of the host population therefore dominates every planet size, and it dominates more completely for the smaller planets.

Because bright WD hosts are rare, we caution that our prediction relies on small-number statistics. Only $\approx10$ of the $16{,}340$ in-field WDs contribute a per-host $N_\mathrm{eff}^\mathrm{PC}$ above $10^{-3}$ for Jupiter-sized planets. These few favorable hosts essentially account for the entire period-averaged yield of $\overline{N}_\mathrm{eff}^\mathrm{PC}\approx0.35$. The gap at $J_\mathrm{AB} = 19$--$20$ is a direct consequence of this sparsity. The $18$--$19$ bin is dominated by a single $\approx28{,}000$\,K host and the $20$--$21$ bin by a single $\approx20{,}000$\,K host, while the $19$--$20$ bin contains only cooler WDs ($\lesssim16{,}000$\,K) and therefore contributes almost nothing. 

Because of these small number statistics, the $N_\mathrm{eff}^\mathrm{PC}$ results above should be interpreted with caution, since the numbers may fluctuate depending on the particular realization of the Besan\c{c}on population synthesis. Fortunately, the bright, hot WDs favorable for phase-curve detections are also the easiest to identify, so Roman will reveal how many such WDs the GBTDS fields contain early in the mission, allowing us to assess the realistic phase-curve yield.

\subsubsection{Eccentricity Dependence}
\label{subsubsec:pc_ecc}

\begin{figure}[t]
    \centering
    \includegraphics[width=\columnwidth]{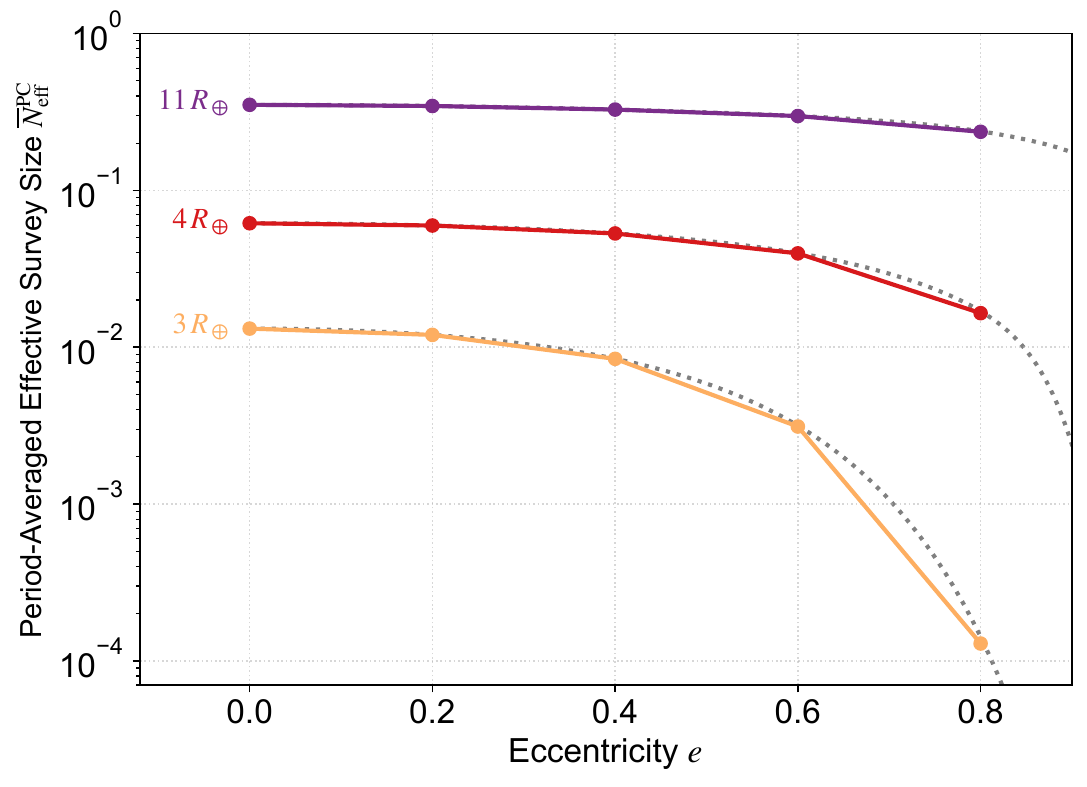}
    \caption{Period-averaged effective survey size $\overline{N}_\mathrm{eff}^\mathrm{PC}$ as a function of eccentricity $e$, color coded by planet size. Gray dotted curves show the predicted yields if eccentricity affects the phase curve only by undersampling the rapid periastron passage and hence lowering the detection SNR.
    }
    \label{fig:pc_eccentricity_dependence}
\end{figure}

Phase curve yield decreases when orbital eccentricity increases, and this decreasing trend is more obvious for smaller planets (Figure~\ref{fig:pc_eccentricity_dependence}). Across the range our grid samples, $e = 0$ to $0.8$, the period-averaged $\overline{N}_\mathrm{eff}^\mathrm{PC}$ of Jupiter-sized planets falls by $\approx33\%$, that of Neptune-sized planets by $\approx73\%$, and that of sub-Neptune-sized planets by two orders of magnitude. This trend is opposite to the transit channel, where a nonzero eccentricity raises $\overline{N}_\mathrm{eff}$ because of transit geometry (Section~\ref{subsubsec:ecc_dependence}).

The decreasing trend in Figure \ref{fig:pc_eccentricity_dependence} is due to a sampling effect rather than geometry or a drop in signal strength. Our energy balance model sets the dayside temperature $T_{\rm day}$ from the semi-major axis (Section~\ref{sec:nontransit_model}), which is fixed once the orbital period is fixed, so an eccentric planet in our grid has the same dayside brightness as a circular one and therefore the same phase curve peak-to-trough amplitude. What matters is how the signal changes over time. A circular orbit produces a clean sinusoidal modulation that Roman can sample uniformly. On an eccentric orbit, however, the planet moves faster near periastron, leading to undersampling where the phase-curve signal changes fastest. As a result of this undersampling, the detection SNR drops as $e$ increases (Figure~\ref{fig:pc_eccentricity_dependence}, gray dotted curve).

However, we note two caveats about how we handle eccentricity in our simulations. First, in reality, the dayside temperature $T_{\rm day}$ depends on eccentricity and the argument of periastron $\omega$, because it tracks the instantaneous planet-star separation rather than the semi-major axis. At secondary eclipse, for example, the separation is $r = a(1-e^2)/(1-e\sin\omega)$ \citep{cowan_statistics_2011}. The second caveat is that planets on eccentric orbits will be heated by tidal deformation, leading to higher emission temperatures than predicted by the energy balance model. The tidal heat flux has a very sensitive dependence on both eccentricity and semi-major axis ($\propto e^2 a^{-15/2}$, see \citealt{driscoll_tidal_2015}), which is not negligible for close-in, eccentric planets \citep{lin_persistent_2026}.

\subsection{Phase Curve Detections Contribute Marginally to Transit Detections}
\label{subsec:complementarity}

The transit and phase curve channels are highly complementary, because they access two almost mutually exclusive populations that overlap only in the rare event of extremely grazing transits. Therefore, we can add the effective survey sizes from both channels together to calculate the total yield of Roman GBTDS at a given planet size: $\overline{N}_\mathrm{eff}^\mathrm{total} \approx \overline{N}_\mathrm{eff}^\mathrm{transit} + \overline{N}_\mathrm{eff}^\mathrm{PC}$.

The phase curve does add to the transit yield, but only marginally. The phase curve channel contributes $\overline{N}_\mathrm{eff}^\mathrm{PC}\approx0.35$ (Table~\ref{tab:pc_yield}) on top of the transit yield of $\overline{N}_\mathrm{eff}^\mathrm{transit}\approx8.01$ (Table~\ref{tab:transit_yield}) for Jupiter-sized planets, raising the combined giant planet yield to $\overline{N}_\mathrm{eff}^\mathrm{total}\approx8.36$, a $\approx4.4\%$ increase. The boost is smaller for Neptune-sized planets, where $\overline{N}_\mathrm{eff}^\mathrm{PC}\approx0.06$ adds to $\overline{N}_\mathrm{eff}^\mathrm{transit}\approx2.60$ for a combined $\approx2.66$ (a $\approx2.4\%$ increase), and even smaller for sub-Neptune-sized planets, where $\overline{N}_\mathrm{eff}^\mathrm{PC}\approx0.01$ adds to $\overline{N}_\mathrm{eff}^\mathrm{transit}\approx1.80$ for a combined $\approx1.81$ (a $\approx0.7\%$ increase). Super-Earth-sized and smaller planets have zero phase curve yield and therefore contribute nothing. The phase curve is therefore a marginal complement to the transit channel in the GBTDS fields rather than a competitive detection method on its own.

\subsection{Phase Curve Yield Framework}
\label{subsec:pc_framework}

Here, we describe in detail how we compute the phase curve effective survey size $\overline{N}_\mathrm{eff}^\mathrm{PC}$ and the corresponding occurrence rate $\overline{\eta}_1$ reported above. The framework is similar to the transit yield calculation (Section~\ref{subsec:yield_framework}), except that the two channels have different geometries and the phase curve detection probability is integrated over orbital inclination $i$.

Similar to the transit channel, we map each of the $16{,}340$ in-field WDs to its nearest grid bin in $(J_\mathrm{AB}, T_\mathrm{eff}, M_\mathrm{WD})$. 
The phase curve effective survey size at a given planet radius and orbital period is the sum of the per-WD detection probabilities,
\begin{equation} \label{eq:neff_pc}
   N_\mathrm{eff}^\mathrm{PC}(R_p, P) = \sum_j \langle P_\mathrm{det} \rangle_j(R_p, P).
\end{equation}
The expected number of detections for an underlying WD exoplanet occurrence rate $\eta_\mathrm{WD}$ is $N_\mathrm{det} = \eta_\mathrm{WD}\, N_\mathrm{eff}^\mathrm{PC}$, and the occurrence rate for one expected detection is $\eta_1 = 1/N_\mathrm{eff}^\mathrm{PC}$. We average $N_\mathrm{eff}^\mathrm{PC}(R_p, P)$ over the same log-uniform period prior using Equation (\ref{eq:neff_int}) to obtain $\overline{N}_\mathrm{eff}^\mathrm{PC}(R_p)$ and the corresponding $\overline{\eta}_1$.

Planetary systems are randomly oriented, so the yield calculation requires $P_\mathrm{det}$ at every inclination, but simulating a dense inclination grid is computationally too costly. We therefore simulate only two inclinations ($i = 20^\circ$ and $80^\circ$) and reconstruct the rest analytically. This is possible because the fraction of the planet's dayside visible to the observer varies as $\sin i$ over an orbit (Section~\ref{sec:nontransit_model}), so the modulation, and hence the expected SNR, depends on the same factor,
\begin{equation} \label{eq:snr_sini}
   \mathrm{SNR}_{\mathrm{exp},j}(i) = \mathrm{SNR}_{\mathrm{exp},j}(i_0) \frac{\sin i}{\sin i_0}.
\end{equation}
A single simulated inclination therefore anchors the whole $i$ axis, and we adopt $i_0 = 80^\circ$, which is nearly edge-on but does not transit. The other inclination grid point ($i = 20^\circ$) serves as a validation point.

Now that $\mathrm{SNR}_\mathrm{exp}(i)$ is known at every inclination, we can evaluate the detection probability. As in the transit channel, $P_\mathrm{det}$ is the probability that a planet at a given grid point can be detected at $>7\sigma$. 
For the phase curve channel, the detection probability for a given orbital inclination $i$ is
\begin{equation} \label{eq:pdet_phi}
   P_\mathrm{det}(i) = \Phi\!\left(\mathrm{SNR}_\mathrm{exp}(i) - 7\right),
\end{equation}
where $\Phi$ is the standard normal cumulative distribution function. Assuming that planetary systems are isotropically oriented, we integrate $P_\mathrm{det}(i)$ over the inclination angle to obtain the orientation-averaged detection probability,
\begin{equation} \label{eq:pdet_marg}
   \langle P_\mathrm{det}\rangle_j = \int_0^{\pi/2} \Phi\!\left(\mathrm{SNR}_{\mathrm{exp},j}(i) - 7\right)\,\sin i\,\mathrm{d}i,
\end{equation}
where the index $j$ denotes a WD system, specified by all grid parameters other than inclination (planet radius, orbital period, and so on).

\section{Discussion} \label{sec:discussion}

In this section, we discuss the implications and limitations of our predictions. We first compare our yields and null-detection limits with existing WD exoplanet surveys and predictions (Section~\ref{subsec:survey_comparison}), and then discuss the astrophysical false positives that a Roman WD transit search must take into consideration, especially in the crowded bulge fields (Section~\ref{subsec:false_positives}). We then consider sub-exposure photometry, which would recover the transit shapes that the GBTDS cadence undersamples (Section~\ref{subsec:subcadence}), and a phase curve search at high Galactic latitude, where source blending is far less severe (Section~\ref{subsec:high_latitude}). We close with limitations of this work and directions for future work (Section~\ref{subsec:limitations}).

\subsection{Comparison with Existing Predictions and Occurrence Rate Limits}
\label{subsec:survey_comparison}

Roman GBTDS is predicted to detect significantly more transiting planets around main-sequence hosts than our yields reported above. By combining the Kepler occurrence rates with a synthetic stellar population, \cite{wilson_transiting_2023} predicted that Roman GBTDS will detect between $60{,}000$ and $200{,}000$ transiting planets, of which roughly $90\%$ are giants ($R_p > 4\, R_\oplus$). \cite{tamburo_predicting_2023} predicted $1347^{+208}_{-124}$ small ($0.5$--$4.0\,R_\oplus$) transiting planets around mid-M dwarfs and ultracool dwarfs. Our period-averaged WD exoplanet yields are three to four orders of magnitude smaller at every planet size, with $\overline{N}_\mathrm{eff}\approx8$ for Jupiter-sized planets and $\overline{N}_\mathrm{eff}\lesssim2.6$ for sizes between $0.5$ and $4\, R_\oplus$ (Section~\ref{subsec:results_yield}). This contrast can be explained by two factors. First, neither earlier forecast includes WD hosts. \cite{wilson_transiting_2023} included only main sequence and subgiant hosts, while \cite{tamburo_predicting_2023} considered only M3--T9 hosts. Secondly, $\overline{N}_\mathrm{eff}$ is not the same quantity as the numbers they reported. Both earlier works quoted an expected number of detections, $N_\mathrm{det} = \eta\, N_\mathrm{eff}$, convolving a measured occurrence rate with their effective survey size. However, we report $\overline{N}_\mathrm{eff}$ alone because the WD exoplanet occurrence rate $\eta_\mathrm{WD}$ is unconstrained.

The primary factor limiting our predicted yield is host brightness. In contrast to the main-sequence forecasts, a WD exoplanet survey cannot restrict the sample to bright targets, because such targets are intrinsically rare. The in-field WD population is faint (Figure \ref{fig:besancon_wd_population_histogram}), with only $24$ of the $16{,}340$ in-field WDs falling in the $J_\mathrm{AB} < 21$ range. Therefore, a WD exoplanet search must extend to fainter hosts, where the typical planetary signal is diluted by a source blending factor of $D\sim10^{-2}$--$10^{-3}$ (Figure \ref{fig:source_blending}). Main-sequence forecasts can avoid this constraint. \cite{wilson_transiting_2023} restricted their sample to the $\approx59$ million stars brighter than \texttt{F146} $= 21$, a limit they adopted because it is approximately where crowding becomes the dominant source of photometric uncertainty, and \cite{tamburo_predicting_2023} adopted the same brightness cutoff. This cutoff is a methodological choice, not a physical limit. \cite{wilson_transiting_2023} argued that a transit survey will likely remain successful at least as faint as \texttt{F146} $\approx22$, roughly doubling the number of monitored stars, because the photometric precision remains adequate for planetary detections. The same reasoning supports extending a WD search to the faint hosts that dominate our sample.

If Roman GBTDS detects no transiting WD planets, the resulting occurrence rate upper limits will be weaker than those already available from K2, even though the GBTDS WD sample is $\sim10\times$ larger. Analyzing the null detection from $1148$ WDs observed by K2, \cite{van_sluijs_occurrence_2018} constrained the occurrence rate of hot Jupiters to $\eta_\mathrm{WD} < 1.5\%$ and habitable zone Earth-sized planets to $< 28\%$. Our period-averaged null-detection limits are $< 28\%$ for Jupiter-sized and $< 88\%$ for Earth-sized planets (Section~\ref{subsec:null_result}), and even our most favorable period-resolved values, $< 6\%$ for Jupiter-sized planets just outside the Roche distance and $< 55\%$ for Earth-sized planets in the shortest period bin, are not tighter than the K2 limits. Source blending is the reason why Roman GBTDS cannot place a tighter constraint. Without source blending, a null detection by GBTDS would constrain Jupiter-sized planets to $< 1.7\%$ (Section~\ref{subsec:null_result}), comparable to the K2 hot Jupiter limit. This Roman GBTDS value, however, is a period-averaged limit spanning the full period range. In contrast, the K2 hot Jupiter constraint applies only to the narrow, close-in region within $a < 0.005$ AU. Thus, the unblended Roman GBTDS limit is arguably a stronger constraint than the K2 limit.

\subsection{False Positives}
\label{subsec:false_positives}

Any exoplanet survey must distinguish astrophysical false positives from genuine planetary signals. Here, we discuss the following false positives that may affect Roman GBTDS: a stellar or substellar companion eclipsing the target WD (Section~\ref{subsubsec:fp_eb}), a diluted eclipsing binary in the background (Section~\ref{subsubsec:fp_background}), disintegrating planetesimals transiting WDs (Section~\ref{subsubsec:fp_debris}), WD pulsations and rotational modulations (Section~\ref{subsubsec:fp_pulsation}), and other photometric variability including Doppler beaming, tidal ellipsoidal variation, and atmospheric modulation from a stellar companion (Section~\ref{subsubsec:fp_beaming}).

\subsubsection{Eclipsing Stellar and Substellar Companions} \label{subsubsec:fp_eb}

WDs are frequently found in binary systems, making eclipsing companions a major source of false positives in any transit survey. Analyses of nearby WDs concluded that roughly 25\% of WDs have binary stellar companions \citep{farihi_low-luminosity_2005, toonen_binarity_2017}. This fraction is lower than the $\sim50\%$ binary fraction of main-sequence stars, but still substantial. The overwhelming majority of WD companions are M dwarfs, while L dwarf, T dwarf, and brown dwarf companions are much less common. \cite{steele_white_2011} concluded that the unresolved WD plus L dwarf, WD plus T dwarf, and WD plus brown dwarf binary fractions are $\geq0.4\pm0.3\%$, $\geq0.2\%$, and $\geq0.5\pm0.3\%$, respectively. Therefore, a WD in the GBTDS fields has a non-negligible chance of hosting a stellar or substellar companion (for convenience, ``substellar'' here refers to very low mass L/T dwarfs and brown dwarfs, but excluding planets), making it necessary to address the false positive induced by eclipsing binaries.

For a transit search, the fraction of companions in orbits close enough to eclipse is more relevant than the total binary fraction. A large fraction of white dwarf plus main sequence (WDMS) binaries are close-in systems that have survived a common envelope phase. From a radial velocity survey of SDSS WDMS binaries, \cite{nebot_gomez-moran_post_2011} found that $21$--$24\%$ of systems have undergone common envelope evolution, with an orbital period distribution that peaks at $\sim10.3$\,h and extends to a few days, partially overlapping with the period range that is most sensitive to transiting exoplanets (Figure \ref{fig:expected_det_vs_eta_WD}b).
Observations have revealed some eclipsing WD systems. \cite{parsons_eclipsing_2013} identified $29$ eclipsing WDMS binaries among $835$ spectroscopically confirmed SDSS systems, and a more recent follow-up found two eclipsing WD substellar companions of almost planetary mass ($\approx19$--$22\, M_\mathrm{Jup}$), which may be misidentified as planets without detailed characterization \citep{parsons_two_2025}.

Examining transit depths can partially resolve eclipsing-binary false positives. If the odd- and even-numbered transits have distinct depths, the source is an eclipsing binary with different brightness. An eclipsing binary with nearly equal brightness can also be ruled out if the odd- and even-numbered transits are both deeper than $50\%$, because the summed depth of the primary and secondary eclipses cannot exceed $100\%$ \citep{vanderburg_giant_2020}. However, shallow planetary transits ($<50\%$) remain indistinguishable from equal-brightness eclipsing binaries. Assuming a typical WD with $M_\mathrm{WD} = 0.6\,M_\odot$ and $R_\mathrm{WD} = 0.0128\,R_\odot$ \citep{kozakis_uv_2018}, a geometric calculation gives the fraction of transits deeper than $50\%$ as $\approx20\%$ for Earth-sized planets, rising to $\approx54\%$, $66\%$, $73\%$, and $89\%$ for super-Earth-, sub-Neptune-, Neptune-, and Jupiter-sized planets, respectively. Therefore, for giant planets that will dominate the Roman GBTDS yield, a small but non-negligible fraction of transits are shallow, so we must consider other mitigation approaches.

One such mitigation approach exploits Roman's infrared capability. Excess near-infrared emission is a standard signature for WDMS or WD plus substellar companion binaries \citep{farihi_low-luminosity_2005, steele_white_2011}. Thanks to the near-infrared coverage of the \texttt{F146} band from $0.93$--$2\,\mu$m, this signature may be accessible. For a small number of favorable candidate systems identified during the survey, follow-up observations with JWST may confirm the planetary mass of the companion, as was done for WD~1856+534~b \citep{limbach_thermal_2025, macdonald_aerosols_2026}.

The out-of-transit light curve offers another way to rule out eclipsing-binary false positives. A stellar or substellar companion produces periodic Doppler beaming, tidal ellipsoidal, and atmospheric modulations (see details in Section~\ref{subsubsec:fp_beaming}), which have distinct characteristics from planetary phase curves. We compare the amplitudes of binary and planetary signals using a representative orbital configuration: a typical $0.6\, M_\odot$ WD with an M3V companion, common for WDMS systems \citep{farihi_low-luminosity_2005}, on a $10.3$\,h orbit \citep{nebot_gomez-moran_post_2011}, versus the same WD with a Jupiter-sized planet on the same orbit. Using the analytic expressions of \cite{shporer_astrophysics_2017}, we find that the stellar companion signal is dominated by Doppler beaming, with a peak-to-trough amplitude of $\approx0.4\%$ that is nearly independent of WD temperature. At the same time, its ellipsoidal and atmospheric components are negligible (both $\approx0.01\%$). The planetary amplitude, in contrast, depends strongly on WD $T_{\rm eff}$, rising from $\approx0.03\%$ for an $8{,}000$\, K WD to $\approx1.3\%$ for a $20{,}000$\, K WD. Therefore, there is an ambiguous parameter space where the $\approx0.4\%$ binary companion signal and the $\approx0.03$--1.3\% planetary signal cross over. Whether a phase curve can rule out the eclipsing-binary false positive therefore depends on the WD temperature and must be assessed individually for each system.

\subsubsection{Background Eclipsing Binaries} \label{subsubsec:fp_background}

Eclipsing binaries in the background whose light is diluted by the foreground target can mimic a planetary transit signal \citep{brown_expected_2003}. In the extremely crowded GBTDS fields, this effect is expected to be a dominant source of false positives.

Mature pipelines developed for past and ongoing transit surveys readily handle background eclipsing binaries. Kepler and TESS data validation pipelines apply a suite of light curve diagnostics to distinguish background eclipsing binaries from genuine planets \citep{coughlin_planetary_2016, thompson_planetary_2018, twicken_kepler_2018}. One canonical way to identify a blended background binary is to measure the photometric centroid in and out of transit \citep{bryson_identification_2013, twicken_kepler_2018}, since a transit originating from a background source shifts the centroid away from the target. \cite{wilson_transiting_2023} argued that Roman is very well suited for transit centroid analysis, because its high spatial resolution and stable point spread function (PSF) enable precise single-exposure astrometry. Even in the pessimistic scenario where the background binary and the target are perfectly aligned, the relative proper motion between the two systems over the multi-year GBTDS baseline will ultimately separate them.

\subsubsection{Transiting Disintegrating Planetesimals}
\label{subsubsec:fp_debris}

Another false positive arises from disintegrating planetesimals transiting a WD. A rocky body orbiting too close to a WD can be tidally disrupted into debris with a comet-like dusty tail, which occults the WD and produces light curve dips that may be misidentified as planetary transits. Transiting disintegrating planetesimals have been observed around a handful of WDs, including WD~1145+017, with debris on $4.5$--$4.9$\,h orbits \citep{vanderburg_disintegrating_2015}, and ZTF~J0328-1219, with debris on $9.9$ and $11.2$\,h orbits \citep{vanderbosch_recurring_2021}.

Planetary debris transits have peculiar shapes and vary over time, so they can be distinguished from transits of intact planets. Disintegrating planetesimals produce asymmetric dips, with sharp ingresses and gradual egresses from the trailing dusty tail, and such dips vary in depth and shape from night to night \citep{vanderburg_disintegrating_2015, gansicke_high-speed_2016, rappaport_drifting_2016}. A transit analysis pipeline can readily reject such irregular signals.

Nevertheless, disintegrating planetesimals orbiting WDs are scientifically interesting. Transiting debris directly probes the tidal disruption process and, when combined with WD pollution observations, can reveal the bulk composition of the rocky materials that once formed those planetesimals \citep{vanderbosch_recurring_2021}. Roman GBTDS may reveal more such disintegrating planetesimals and even capture tidal disintegration in real time.

\subsubsection{White Dwarf Pulsations and Rotational Modulations}
\label{subsubsec:fp_pulsation}

Most WDs go through a stage of pulsational instability during their evolution \citep{corsico_pulsating_2019}. During this stage, WDs become multi-periodic pulsating variable stars, producing brightness changes that may be misidentified as planetary phase curves. The most common type of pulsating WDs are the ZZ~Ceti stars. With $10{,}400 \lesssim T_{\rm eff} \lesssim 12{,}400$\,K, $7.5 \lesssim \log g \lesssim 9.1$, and nearly pure hydrogen atmospheres \citep{fontaine_pulsating_2008, corsico_pulsating_2019}, ZZ~Ceti stars coincide with common characteristics of Besan\c{c}on WDs (Figure~\ref{fig:besancon_wd_population_histogram}). The hotter V777~Her stars ($22{,}400 \lesssim T_{\rm eff} \lesssim 32{,}000$\, K), while less common, fall in the WD temperature range that favors phase curve detections (Section \ref{subsubsec:pc_period_dependence}). In addition, WDs can display periodic photometric variability produced by stellar rotation combined with magnetic or other surface inhomogeneities \citep[e.g.,][]{kilic_dark_2015, reding_isolated_2020}.

In practice, pulsational signals are usually easy to distinguish from planetary phase curves, but rotational modulations are harder to identify. Pulsating WDs are typically multi-periodic, while a planetary phase curve has strictly one period, so the presence of several incommensurate periods implies stellar pulsation. Rotational modulation, however, is mono-periodic. Very rapid rotators, with periods as short as a few minutes or even tens of seconds \citep[e.g.,][]{caiazzo_highly_2021, kilic_isolated_2021}, can be excluded because no planet can survive on such orbits. Isolated WDs, however, typically rotate with periods of $0.5$--$2.2$\,d \citep{hermes_white_2017}, which overlap the orbital periods of the planets Roman GBTDS is most sensitive to, so a mono-periodic modulation in this range is ambiguous and requires detailed follow-up observations.

\subsubsection{Other Photometric Modulations Induced by a Companion}
\label{subsubsec:fp_beaming}

A stellar or substellar companion produces periodic photometric modulations even when it does not transit, and such modulations can be confused with planetary phase curves. There are three processes leading to photometric modulations: Doppler beaming and atmospheric modulation, which vary at the binary orbital period, and tidal ellipsoidal distortion, which varies at half the orbital period \citep{faigler_photometric_2011, shporer_astrophysics_2017}. These modulations are routinely used to detect non-transiting companions, and TESS observations have confirmed them for a handful of systems, including four with significant ellipsoidal distortion and three with Doppler beaming \citep{wong_systematic_2020}.

The WD host, however, makes these false positives easier to identify than a main-sequence host. The ellipsoidal component is particularly diagnostic. Its amplitude scales as $A_\mathrm{ellip} \propto 1/\rho_\star$, inversely with the bulk density of the distorted body \citep{shporer_astrophysics_2017}. A WD is far too dense to be distorted to any detectable degree, and a planet is not self-luminous, so its distortion contributes minimally to the combined flux. Therefore, a significant ellipsoidal signal must originate from a self-luminous stellar or substellar companion. Because the ellipsoidal signal is the only one of the three that varies at half the orbital period, we can identify this false positive if two periods differing by a factor of two are detected in the phase curve.

These modulations play opposite roles in the two detection channels. In the phase curve channel, they act as false positives. In the transit channel, however, the same modulations become useful diagnostics. As noted in Section~\ref{subsubsec:fp_eb}, a transiting stellar or substellar companion can be identified from these photometric modulations in the out-of-transit light curve, especially when the WD $T_{\rm eff}$ or orbital configuration does not favor a hot giant planet.

\subsection{Sub-Exposure Data}
\label{subsec:subcadence}

Roman GBTDS undersamples WD exoplanet transits. Roman will take a single 66-s exposure of each field every 12.1 minutes. Because this interval is much longer than the exposure itself, \cite{wilson_transiting_2023} noted that the transit shape is undersampled, degrading transit-detection sensitivity most for the shortest transit durations. Short transit durations, however, are typical for WD systems because of their small stellar size. WD~1856+534\,b has a transit duration of only $\approx8$ minutes \citep{vanderburg_giant_2020}.

This undersampling effect can be partially mitigated by analyzing sub-exposure data. The Roman WFI detectors are read out non-destructively, so the default read-out strategy is to ``sample up the ramp,'' recording the accumulated voltage of each pixel every 3.04 seconds \citep{wilson_transiting_2023, tovar_mendoza_enabling_2023}. Variations faster than the exposure time are therefore not lost and can, in principle, be recovered from the individual reads rather than collapsed into a single flux value per exposure. \cite{tovar_mendoza_enabling_2023} proposed this approach for stellar flare detections using GBTDS, which, like WD transits, are short-duration events vulnerable to undersampling.

For WD exoplanets, sub-exposure photometry would constrain transit shapes better than full exposures do. Resolving the rapid ingress and egress of a WD exoplanet transit informs the impact parameter and limb darkening. In addition, the transit shape helps reject false positives such as the asymmetric, time-variable dips of disintegrating planetesimals (Section~\ref{subsubsec:fp_debris}).

\subsection{High Latitude Phase Curve Search}
\label{subsec:high_latitude}

A phase curve search is plausibly more suitable for Roman's HLTDS than for the GBTDS. As demonstrated in Section \ref{sec:results_nontransit}, phase curve observations by Roman GBTDS will yield at best a marginal detection of WD exoplanets. The GBTDS phase curve search is limited by source blending in the crowded Galactic bulge and center fields and by its reliance on a very small number of bright hosts. Both limitations will ease at high Galactic latitudes. Fields away from the Galactic plane are far less crowded, and WDs observed there are nearby rather than at kpc bulge distances, so finding bright, undiluted sources is more feasible.

Roman's HLTDS will image roughly $25\,\mathrm{deg}^2$ of sky at high Galactic latitudes at a 5-day cadence over two years \citep{zasowski_roman_2025}. Such a long cadence will prevent transit detections, but it is not a serious limitation for a phase-curve search, which targets continuous modulation. Over two years, a 5-day cadence accumulates enough epochs to sample the full orbits of all WD planets except those with the longest periods, whose low $T_{\rm day}$ places them below detectability regardless of cadence.

A rough estimate suggests that the HLTDS will monitor fewer WDs than the GBTDS, but of better quality. \cite{fantin_white_2020} predicted that Roman's High-Latitude Wide-Area Survey (HLWAS) will detect $735{,}401$ WDs across a sky area of $2213\,\mathrm{deg}^2$, or $\approx332$ WDs per $\mathrm{deg}^2$ if WDs are distributed uniformly. Given this density, the roughly $25\,\mathrm{deg}^2$ HLTDS footprint would contain $\approx8{,}300$ WDs, about half the $16{,}340$ in the GBTDS fields, even though it covers roughly $15\times$ more sky than the $\approx1.7\,\mathrm{deg}^2$ GBTDS footprint. The WDs at higher latitudes, however, would be far less diluted by background sources. One important caveat is that the HLTDS filter set does not include \texttt{F146}, so the yields reported here cannot be transferred directly to the high-latitude fields. We leave a quantitative prediction for future investigations.

\subsection{Limitations and Future Work}
\label{subsec:limitations}

The synthetic WD population is the single largest source of systematic uncertainty in our yields. We adopt the Besan\c{c}on model for our yield simulation because it provides a Galactic model in which WDs are treated as a separate class with consistent dynamics and built-in dust extinction (Section~\ref{subsec:population}). To maintain self-consistency, Besan\c{c}on does not allow users to adjust many parameters, so quantifying uncertainties in individual model parameters will require a different population model. A useful cross-comparison would be to regenerate our population with \textsc{synthpop} \citep{kluter_synthpop_2025}, a modular population synthesis framework in which the density profile, initial mass function, age and metallicity distributions, isochrones, and extinction treatment are all specified independently. \textsc{synthpop} also provides Roman bands directly, which would remove the need for $J$ band as proxy for \texttt{F146}, although its WD properties are derived from isochrones and would require validation before being adopted for our purpose.

The predictions in this work rely on simulated light curves. Converting simulations into actual detections will require a pipeline that operates on real Roman data products. The Microlensing Science Operations System (MSOS)\footnote{\url{https://roman-docs.ipac.caltech.edu/data-handbook/roman-wfi-data-pipelines/galactic-bulge-survey-pipelines}} produces Level-3 and Level-4 products, including a fiducial catalog of all detected objects and a per-epoch light curve catalog derived from PSF fitting and difference imaging analysis, which are both central to a planet search. A WD exoplanet survey can therefore begin from these high-level products. WD hosts would first be identified photometrically, using the high-cadence \texttt{F146} imaging together with the \texttt{F087} and \texttt{F213} observations obtained every 6 hours and photometric snapshots in five additional bands, three epochs per season. Preselecting WD hosts is essential because the light curve catalog will produce 3 TB of data daily.

We need to develop a transit and phase-curve search pipeline to find planets in the light-curve products. The MSOS pipeline is designed for microlensing events rather than transits, and standard transit search methods such as box least squares \citep{kovacs_box-fitting_2002} and transit least squares \citep{hippke_optimized_2019} typically assume planets are much smaller than their hosts. Hence, the deep (or even total) and undersampled transits of WD systems require a dedicated search pipeline. Once candidates are identified, the fitting and vetting stages can build on mature community tools, such as \texttt{allesfitter} \citep{gunther_allesfitter_2021}, which we already use in our injection-recovery test (Section~\ref{subsec:injection}), \texttt{juliet} \citep{espinoza_juliet_2019}, \texttt{exoplanet} \citep{foreman-mackey_exoplanet_2024}, and \texttt{EXOFASTv2} \citep{eastman_exofastv2_2019}. In future work, we plan to develop a search pipeline optimized for WD hosts under the approved Roman Cycle 1 General Investigator program \#19060,\footnote{\url{https://roman.ipac.caltech.edu/cycle1-approved-programs/19060}} couple it to these tools, and release the resulting candidate catalog as a Level-5 community data product.

\section{Conclusions} \label{sec:conclusions}

The Roman Space Telescope will monitor $\sim16{,}000$ white dwarfs in the Galactic bulge and center at 12.1-minute cadence during six high-cadence GBTDS seasons over the 5-year mission baseline. This combination of short cadence and long baseline provides the ideal configuration to capture the brief transits and weak phase-curve amplitudes of WD planetary systems. In this work, we predict how many WD exoplanets the GBTDS will detect over its first three high-cadence seasons, through both the transiting and the non-transiting (phase curve) channels. We build a library of WD spectral energy distributions, pass them through a \texttt{Pandeia} per-exposure noise model to obtain the Roman \texttt{F146} photometric precision, and run a Monte Carlo SNR and detection probability simulation over a large grid of planet radius, orbital period, and host properties across the $\approx436$-day baseline. The SNR and detection probability grid is validated by \texttt{allesfitter} injection-recovery tests. We then fold this grid over a Besan\c{c}on synthetic population of $16{,}340$ WDs in the GBTDS fields, remove unphysical planets within the Roche limit, and correct for source blending in the crowded bulge fields. The results are reported as the effective survey size $N_\mathrm{eff}(R_p, P)$, the number of detections expected if every in-field WD hosts one planet of radius $R_p$ at orbital period $P$, and its period-averaged counterpart $\overline{N}_\mathrm{eff}$, marginalized over a log-uniform period prior, together with $\overline{\eta}_1 = 1/\overline{N}_\mathrm{eff}$, the occurrence rate required for GBTDS to expect one detection.

Our main findings for the transit channel are as follows.

\begin{itemize}
    \item Jupiter-sized planets dominate the transit yield, with a period-averaged effective survey size of $\overline{N}_\mathrm{eff}\approx8.01$, followed by $2.60$ for Neptune-sized, $1.80$ for sub-Neptune-sized, $1.06$ for super-Earth-sized, $0.43$ for Earth-sized, and $0.13$ for Mars-sized planets. The corresponding required occurrence rates are $\overline{\eta}_1\approx12\%$, $38\%$, $55\%$, $94\%$, $230\%$, and $771\%$, respectively.
    \item In the crowded bulge fields, dilution from background sources removes $\approx95\%$ of the Jupiter-sized yield and $\approx97$--$98\%$ for Neptune-sized and smaller planets, because the in-field WDs are faint (median $J_\mathrm{AB}\approx25.4$) and are typically diluted by a factor of $D\sim10^{-2}$--$10^{-3}$.
    \item The effective survey size concentrates at short period. $N_\mathrm{eff}$ peaks just outside the Roche distance, reaching $\approx47$ at $P\approx0.5$\,d for Jupiter-sized planets ($\eta_1\approx2.1\%$), which is roughly $6\times$ greater than the period-averaged value. The unknown WD exoplanet period distribution therefore introduces a large uncertainty in the yield.
    \item The giant planet yield is dominated by WDs of intermediate brightness, peaking in the $J_\mathrm{AB}=23$--24 bin. This reflects a balance between per-WD detectability, which favors bright hosts, and host number, which rises toward the faint end.
    \item A nonzero eccentricity raises the transit yield, increasing $\overline{N}_\mathrm{eff}$ by a factor of $\approx2.6$ at $e=0.8$, nearly uniform across planet size, because the geometric transit probability scales as $1/(1-e^2)$. The $e=0$ values are therefore conservative lower bounds.
    \item A null result will still be informative. Assuming a log-uniform period prior, a null detection would constrain the WD exoplanet occurrence rate to $\eta_\mathrm{WD}<28\%$ for Jupiter-sized planets, $<57\%$ for Neptune-sized, $<66\%$ for sub-Neptune-sized, $<77\%$ for super-Earth-sized, $<88\%$ for Earth-sized, and $<93\%$ for Mars-sized planets, all at $95\%$ confidence. With period resolution, the constraints tighten considerably where the survey is most sensitive, reaching $<6\%$ for Jupiter-sized planets just outside the Roche distance and $<55\%$ for Earth-sized planets at the shortest sampled period. Roman GBTDS will provide the first-ever WD exoplanet occurrence rate limits for the Galactic bulge and center.
\end{itemize}

Phase curve offers a complementary way to search for planets that do not transit. Our main findings for the phase curve channel are as follows.

\begin{itemize}
    \item Phase curve observations by Roman GBTDS will provide at best a marginal detection of WD exoplanets, with the limited sensitivity confined to a narrow parameter space of close-in giant planets. The phase curve effective survey size is $\overline{N}_\mathrm{eff}^\mathrm{PC}\approx0.35$ for Jupiter-sized planets ($\overline{\eta}_1\approx285\%$), $0.062$ for Neptune-sized, $0.013$ for sub-Neptune-sized, and $\approx0$ for super-Earth-sized and smaller planets.
    \item Phase curve detection is challenging because the signal is intrinsically weak. Even for the most favorable system configuration, a hot ($20{,}000$\,K), low mass ($0.4\,M_\odot$), bright ($J_\mathrm{AB}=18$) WD hosting a Jupiter-sized planet at $P = 12.2$\,hr, the phase curve amplitude reaches only $\approx2\%$ from peak to trough, $\approx50\times$ smaller than the total occultation the same planet would produce if it transits.
    \item Similar to the transit channel, source blending in the crowded bulge fields is a major obstacle, removing $\approx97\%$ of $\overline{N}_\mathrm{eff}^\mathrm{PC}$ for Jupiter-sized planets.
    \item Phase curve detectability is confined to close-in orbits around hot hosts. The effective survey size of Jupiter-sized planets peaks at $N_\mathrm{eff}^\mathrm{PC}\approx4.2$ at $P\approx0.5$\,d, immediately outside the Roche distance, then falls by more than three orders of magnitude by $P\approx1.3$\,d. The $3{,}104$ in-field WDs hotter than $\approx16{,}000$\,K, $\approx19\%$ of the sample, supply $\approx98\%$ of the predicted yield.
    \item Only bright hosts contribute to the yield. Nearly the entire Jupiter-sized yield comes from three bins, $J_\mathrm{AB} = 18$--19, 20--21, and 21--22, contributing $\approx31\%$, $\approx30\%$, and $\approx38\%$ respectively, while WDs fainter than $J_\mathrm{AB}=22$ supply only $\approx0.05\%$. Smaller planets require brighter hosts: $\approx62\%$ of the Neptune-sized yield and $\approx91\%$ of the sub-Neptune-sized yield come from the single $J_\mathrm{AB}=18$--19 bin.
    \item A nonzero eccentricity lowers the phase curve yield, in contrast to the transit channel. From $e=0$ to $0.8$, $\overline{N}_\mathrm{eff}^\mathrm{PC}$ falls by $\approx33\%$ for Jupiter-sized planets and $\approx73\%$ for Neptune-sized planets, because the rapid periastron passage is undersampled at the survey cadence.
    \item Transit and phase curve channels access nearly mutually exclusive populations, so their effective survey sizes can be summed together. Phase curve detections raise the Jupiter-sized effective survey size from $\overline{N}_\mathrm{eff}\approx8.01$ to $\approx8.36$, a $\approx4.4\%$ increase, and the Neptune-sized value from $\approx2.60$ to $\approx2.66$, a $\approx2.4\%$ increase, making the phase curve a marginal complement to the transit channel in the GBTDS fields rather than a competitive detection method on its own.
\end{itemize}

The white dwarf SED library, synthetic white dwarf catalog, detection-probability grids, and background blending and dust extinction maps used in this work are publicly available on Zenodo \citep{lin_detecting_2026}.

The predictions presented in this work have two main limitations. First, the synthetic WD host population is the largest source of systematic uncertainty. Besan\c{c}on does not allow many of its parameters to be varied independently, so quantifying their effect on the final yield will require simulating the population with a modular framework such as \textsc{synthpop} \citep{kluter_synthpop_2025}. Second, the underlying period distribution of WD planets is unconstrained, and it strongly shapes the yield, as seen in the contrast between $N_{\rm eff}$ and $\overline{N}_{\rm eff}$. Our $N_\mathrm{eff}(R_p, P)$ grid can be refolded with any period distribution so that future constraints will refine the yields reported here.

Converting these predictions into detections will require a special search pipeline built for WD hosts in the Roman dataset. We plan to develop such a pipeline based on the MSOS light curve products, couple it to existing fitting and vetting tools, and release the resulting candidate catalog as a community data product. Beyond the first three seasons, the full six-season baseline will reach planets on wider orbits, and sub-exposure photometry will recover the detailed transit shapes that the 12.1-minute cadence undersamples. Whether Roman GBTDS returns detections or a null result, it will deliver the first population-level constraints on planets orbiting WDs in the Galactic bulge, opening a new observational window onto the fate of planetary systems after post-main-sequence evolution.

\begin{acknowledgments}
Z.L. and T.D. acknowledge support from the McDonnell Center for the Space Sciences at Washington University in St. Louis.
\end{acknowledgments}

\begin{contribution}

Z.L. conceived the study, developed the simulation framework, performed the analysis, and drafted the manuscript. T.D. supervised the project and edited the manuscript. All authors reviewed and approved the final manuscript.


\end{contribution}

%

\software{\texttt{Pandeia} \citep{pontoppidan_pandeia_2016},
          \texttt{pysiaf} \citep{shannon_osborne_spacetelescopepysiaf_2026},
          \texttt{allesfitter} \citep{gunther_allesfitter_2021}
          }






\bibliographystyle{aasjournalv7}
\bibliography{roman_wd}



\end{CJK*}
\end{document}